\pdfoutput=1
\documentclass[aps,reprint,twocolumn,prd,nofootinbib]{revtex4-2}

\usepackage{hyperref}
\hypersetup{colorlinks=true, citecolor=blue}

\usepackage{amsmath,amssymb,amsfonts,mathrsfs}

\usepackage{mathtools}
\usepackage{bm}
\usepackage{graphicx}   
\usepackage[utf8]{inputenc}
\usepackage{latexsym}

\newcommand{\dd}{\mathrm{d}}

\begin{document}

\title{Searching for topological dark matter in LIGO data}

\author{Lavinia Heisenberg}
\email{heisenberg@thphys.uni-heidelberg.de}
\affiliation{Institute for Theoretical Physics, Heidelberg University, Philosophenweg 16
D-69120	Heidelberg,
Germany} 

\author{David Maibach}
\email{d.maibach@thphys.uni-heidelberg.de}
\affiliation{Institute for Theoretical Physics, Heidelberg University, Philosophenweg 16
D-69120	Heidelberg,
Germany} 

\author{Do\u{g}a Veske}
\email{veske@thphys.uni-heidelberg.de}
\affiliation{Institute for Theoretical Physics, Heidelberg University, Philosophenweg 16
D-69120	Heidelberg,
Germany}

%%%%%%%%%%%%%%%%%%%%%%%%%%%%%%%%%%%%%%%%%%%%%%%%%%%%%%%%%%%%%%%%%%%%%%%%%%%%%%%%%%%%%%%%%%%%%%

\begin{abstract}

Gravitational-wave interferometers have been recently proposed as a promising probe in searches for dark matter. These highly sensitive instruments are potentially able to detect the interactions of dark matter with the detector's hardware. In this work, we explore the possibilities of discovering topological dark matter with the LIGO detectors. We analyze domain walls consisting of dark matter passing through the Earth, leaving traces in multiple detectors simultaneously. Considering dark matter interactions with light in the interferometer, and with the beam splitter, we perform the first analysis of topological dark matter with gravitational-wave strain data. We examine whether astrophysically unexpected triggers could be explained by domain-wall passages. We find that all of the binary black hole mergers we analyze favor the binary black hole merger hypothesis rather than the domain-wall hypothesis, with the closest being GW190521. Moreover, we find that some topological dark matter signals can be caught by binary black hole searches. Finally, we find that special types of glitches in the measurement data can inevitably limit the dark matter searches for certain parameters. These results are expected to guide future searches and analyses.

\end{abstract}

%%%%%%%%%%%%%%%%%%%%%%%%%%%%%%%%%%%%%%%%%%%%%%%%%%%%%%%%%%%%%%%%%%%%%%%%%%%%%%%%%%%%%%%%%%%%%%

\maketitle

%%%%%%%%%%%%%%%%%%%%%%%%%%%%%%%%%%%%%%%%%%%%%%%%%%%%%%%%%%%%%%%%%%%%%%%%%%%%%%%%%%%%%%%%%%%%%%

\section{Introduction} \label{intro}
Dark matter (DM) and dark energy (DE) comprise most of the Universe's energy content. However, both are poorly understood, and an elucidation of their nature remains one of the most tantalizing problems in modern physics. While DE is mostly assumed to be a cosmological constant, the nature of DM is investigated using a wide range of theories \cite{Feng_2010, Arbey_2021, khoury2015dark, adams2023axion}. Despite the evidence for DM originating from its gravitational interactions, deciphering its nature necessarily relies on nongravitational interactions between DM and the constituents of the Standard Model of particle physics. However, detecting low-mass DM via particle-like interactions is practically impossible due to the particles' comparably small momenta. Instead, one commonly assumes that the large occupation number of low-mass DM particles must result in wave-like and other coherent signatures. This picture has driven the development of novel experiments borrowing techniques from precision measurements in atomic and optical physics to which the DM field (frequently described as a scalar field) is susceptible \cite{Flambaum_2023, Safronova_2018, Derevianko_2014}. In particular, it has been proposed to use ground-based gravitational-wave (GW) interferometers such as LIGO~\cite{2015ligo}, Virgo~\cite{2015CQGra..32b4001A} or KAGRA~\cite{10.1093/ptep/ptaa125} as precision measurements for effects of DM coupled to the gauge sector of the Standard Model \cite{Jaeckel_2016, PhysRevResearch.1.033187, Khoze_2022, Pospelov_2013, Vermeulen_2021}. Originally constructed for the sole purpose of detecting GWs, these cutting-edge interferometers operate at a level of precision that is able to measure even very low-mass DM effects when an accumulation of such is placed within the pathway of the detector beam. The ensemble of nontrivial DM field values induces local, temporal variations of fundamental physical ``constants" such as the fine-structure constant $\alpha$, the speed of light, and fermion masses via nongravitational interactions with the standard sector. The time-dependent fluctuations of these constants will impact the phase differences of the detector arms if the distribution of DM is split unequally between them. Such a concentrated grouping of DM can arise, for instance, in the form of a topological defect, specifically as a domain wall (DW). A DW consisting of the DM field passing through the detector with a normal vector inclined towards one of its arms would induce the previously mentioned effects. Recent studies have examined potential signatures of DM within interferometer setups \cite{Jaeckel_2016, Pospelov_2013, Vermeulen_2021}. Additionally, investigations into various types of DM have utilized GW data \cite{Vermeulen_2021,Pierce_2018,Guo_2019,Abbott_2022}. Notably, several works have explored DM signatures through GW emissions \cite{fell2023detecting,PhysRevLett.131.091401,PhysRevLett.126.141105,PhysRevD.96.035019,PhysRevLett.125.141101,Maselli_2022,PhysRevLett.122.211301}. However, as of yet, there has been no data analysis specifically targeting topological DM using GW data. The current constraints on the DM field, especially in its localized manifestation as a topological defect, stem from experiments that do not utilize  GW interferometers \cite{Afach_2021,doi:10.1126/sciadv.aau4869,Roberts_2017,PhysRevD.102.115016,Yang_2021,Roberts_2020}.

In this work, we undertake the first investigation with real instrumental data and establish a framework for the exploration of topological DM within data of GW detectors. To do so, we revisit the proposition that DM fields can give rise to stable topological defects (monopoles, cosmological strings, or DWs) \cite{Sikivie_1982, Kawasaki_2015} which significantly contribute to the total energy density attributed to DM.  Our focus lies predominantly on DM configurations forming DW solutions \cite{Avelino_2008, Friedland_2003} which we henceforth refer to as topological DM (TDM). DWs naturally emerge in theories where discrete symmetries are broken, a common occurrence in scenarios beyond the Standard Model in particle physics. For the broken symmetry in question, simple $\mathbb{Z}_2$ symmetries are widely used in the context of two-Higgs-doublet models and hence apply well for DW candidates. Various other discrete symmetries arise in different branches of physics, such as flavor physics, which could potentially give rise to DWs as well. In this article, we study DM consisting of a scalar (spinless, even-parity) field $\phi$ \cite{Pospelov_2013} with weak nonlinear-in-$\phi$ interactions with Standard Model particles. In this form, the free DM Lagrangian can be written as 
\begin{equation}
\label{equ:1}
\mathcal{L}_{\textsc{DW}} = \frac{1}{2} (\partial \phi) ^2 - \frac{2 m_{\text{dark}}^2 f^2}{N_\phi^2}
\sin^2\left(\frac{N_\phi \phi}{2f}\right)\,,
\end{equation}
which has DW solutions 
\begin{equation}
\label{equ:2}
    \phi_{\textsc{TDW}}(x)= \frac{4f}{N_\phi} \arctan(e^{m_{\text{dark}}\cdot x})\,.
\end{equation}
Hence, the Lagrangian \eqref{equ:1} gives rise to a suitable TDM model\footnote{For simplicity, the solution \eqref{equ:2} is adapted to the $1$-dimensional case. We will discuss a realistic $3+1$-dimensional setup below.}. Here, $f$ is the scale of the vacuum energy and $m_{\text{dark}}$ is the mass of a small excitation of $\phi_{\textsc{TDW}}$ around any minimum of the potential. The integer parameter $N_\phi$ establishes a shift symmetry $\phi \rightarrow \phi + \frac{2\pi}{N_\phi}f$ as a consequence of the $\mathbb Z_{N_\phi}$ symmetry of the underlying complex scalar field theory $\Phi(x) = S(x)e^{i\phi(x)/f}$. The thickness $d$ of the DW is determined by the mass as $d\sim 2/m_{\text{dark}}$.

As mentioned before, DWs of thickness $d$ passing through the detector leave discernible imprints that are observable with present instrumentation. Specifically, here we consider LIGO as a potential detection tool. The frequency of DWs traversing the Earth (and thus Earth's detectors) can be constrained via the DM energy density within the vicinity of the Solar System, $\rho_{\text{DM}} \simeq 0.4 \text{ GeV/}\text{cm}^3$ \cite{PhysRevD.98.030001}. Assuming that TDM forms a fractional part of $\rho_{\text{DM}}$, one finds that 
\begin{equation}
    \rho_{\text{TDM}} \sim \sigma/L \leq \rho_{\text{DM}}\,.
\end{equation}
In a cosmological context, we disregard the width of the DW and define $\sigma$ as the mass per unit area, i.e.,
\begin{equation}
    \sigma = \int \dd x\left| \frac{\dd \phi_{TDW}}{dx}\right|^2\,.
\end{equation}
Then, we can define an additional distance parameter describing the average separation of DWs, $D_{\rm DW}$. Using \eqref{equ:2}, and reintroducing the previously omitted squared factors of the reduced Planck's constant $\hbar$ and speed of light $c$, the average separation can be constrained to 
\begin{equation}
    D_{\rm DW}\geq \frac{8 f^2 m_{\text{dark}} }{\hbar^2 c^2 N^2_\phi\cdot  0.4 \text{GeV/}\text{cm}^3}\,.
\end{equation}
Choosing a vacuum scale similar to the Higgs sector, i.e., $250 \text{ GeV}$, $m_{\rm dark}\sim 10^{-12}$ eV, and $N_\phi$ as unity, the resulting lower bound on the average separation scale for TDM is $D_{\rm DW}\gtrsim 3\times 10^{10}$m. Assuming a DW speed of $\sim10^{-3}c$, the latter bound corresponds to $\lesssim 1$ DW crossing per day. Moreover, we can estimate a number density for the DWs based on the lower bound for $D_{\rm DW}$. With the volume of the Solar System's neighborhood being estimated conservatively to be roughly $10^{30}\text{ km}^3$, we find that, given the above constraints, there will be at most $\mathcal{O}(10^7)$ DW per Solar System or, in other words, $n_{\text{DW}}\leq 10^{18} \text{ pc}^{-3}$. As this upper bound does not violate any particle physics constraints, the occurrence of DW crossings could potentially yield multiple events during the duration of LIGO detector runs and, thus, the search for TDM in GW detector data is well motivated. 

This article promotes the hunt for DM in GW interferometer data by showing that existing events and frequently reoccurring types of glitches match the signatures resulting from TDM passing through the Earth. It is structured as follows: In Sec. \ref{dm_model} we present the interaction types between DM and Standard Model particles of interest and the resulting impact on the phase signals in the GW interferometers. We specifically discuss ground-based interferometers with Fabry-P\'erot cavities such as the LIGO, Virgo, and KAGRA detectors. In Sec. \ref{ana} we provide concrete evidence for TDM-like signatures regularly occurring in LIGO data in the form of either signals identified as massive binary black holes (BBH) or glitches. We conclude this article with a discussion and remarks for future works in Sec. \ref{discus}.

%%%%%%%%%%%%%%%%%%%%%%%%%%%%%%%%%%%%%%%%%%%%%%%%%%%%%%%%%%%%%%%%%%%
                    %%
                    %%
%%%%%%%%%%%%%%%%%%%%%%%%%%%%%%%%%%%%%%%%%%%%%%%%%%%%%%%%%%%%%%%%%%%
\section{Dark Matter Modeling} \label{dm_model}
We start by outlining the types of interactions examined in the analysis, outlined in Sec. \ref{ana} of this study. As previously mentioned, our focus lies exclusively on scalar field DM $\phi$ coupling to various Standard Model fields.

\subsection{Interactions with Standard Model particles}
In general, a plethora of possible interactions could be incorporated into the Lagrangian \eqref{equ:1}, constrained solely by current observational and theoretical limitations\footnote{These may encompass, for instance, parity or (Lorentz) symmetry violations of any nature as well as any constraints from particle physics experiments.}. Here, we wish to consider the three most prominently used representatives of a wider range of interaction types (collecting the results of \cite{Khoze_2022, Jaeckel_2016, Pospelov_2013}). To demonstrate the most relevant effects, we start by considering a coupling of the DM field to the canonical photon kinetic term 
\begin{equation}\label{equ:0}
    \mathcal{L}_{int}\supset \frac{1}{4} g_{\text{DW}} \left(\frac{\phi}{f}\right)^2 F_{\mu\nu}F^{\mu\nu} - \sum_{\Psi_i}  \lambda_{\text{DW},i}\left(\frac{\phi}{f}\right)^2 m_{\Psi_i} \Psi_i\bar \Psi_i\,,
\end{equation}
which was complemented by a fermionic mass coupling, i.e., the second term in \eqref{equ:0}. In the latter, the field $\Psi_i$ is an arbitrary fermion species with mass $m_{\Psi_i}$ while $F^{\mu\nu}$ denotes the standard electromagnetic field-strength tensor. This type of interaction term introduces $n+1$ fiducial coupling constants, one for the kinetic term $g_{\text{DW}}$ and $n$ mass couplings $\lambda_{\text{DW},i}$ where each fermion species $i$ receives a unique coupling constant. Note that, as the linear couplings vanish in the case of an underlying $\mathbb Z_2$ symmetry (invariance under $\phi\rightarrow -\phi$), we directly impose a quadratic coupling. Choosing $f$ to be large, we see that higher-order interactions in $\phi/f$ become irrelevant automatically.

\begin{figure}
\centering
  \centering
  \includegraphics[width=1\linewidth]{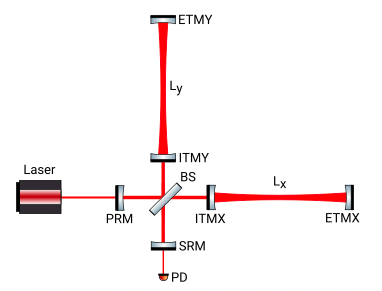}
  \caption{Schematic sketch of a general dual-recycled Fabry-P\'erot-Michelson-type interferometer from \cite{Grote_2019}. The assembly displayed here depicts core features inherent to Advanced LIGO \cite{2015ligo}, Virgo \cite{2015CQGra..32b4001A} and KAGRA \cite{10.1093/ptep/ptaa125}.}
  \label{fig:RecycleFPI}
\end{figure}

The interaction of type \eqref{equ:0} will impact the fine-structure constant due to the presence of a kinetic photon coupling as well as individual fermion masses via the mass couplings. One can easily show that the interaction term results in 
\begin{equation}
    \alpha \rightarrow\frac{\alpha}{1-g_{\textsc{DW}}\frac{\phi^2}{f^2}} , \ m_\Psi\rightarrow \frac{m_\Psi}{1-\lambda_{\textsc{DW}}\frac{\phi^2}{f^2}}\;.
\end{equation}
Alterations of the fine-structure constant and/or the fermion mass induce a change in the optical path length of each interferometer arm via perturbing the Bohr radius and the refraction index of the mirrors and the beam splitter in the instrument (see a sketch in Fig. \ref{fig:RecycleFPI}). 

In general, a time-dependent $\alpha$ and fermion mass lead to time modulations of the spatial extent of solid objects including the test masses of an interferometer. As the Bohr radius depends on both $\alpha$ and $m_\Psi$, a solid body's size change in the adiabatic limit yields
\begin{equation}
    \frac{\delta L }{L} \approx -\frac{\delta \alpha}{\alpha} -\frac{\delta m_\Psi}{m_\Psi}\;.
\end{equation}
The latter equation is to be regarded for all solid objects within the interferometer arms which, over time, are traversed by nontrivial DM field configurations. For the instruments in question, this effect primarily concerns the mirrors and the beam splitter.\footnote{We comment on effects related to the interferometer's test masses in Section \ref{signals_TDM}.} Note here that most ground-based dual-recycled Fabry-P\'erot-Michelson type interferometers (e.g., LIGO, Virgo, KAGRA; see Fig \ref{fig:RecycleFPI}) have additional mirrors (``ITMY" and ``ITMX") increasing the strain sensitivity of the detector with respect to GWs. For the detection of TDM, the additional mirrors do not increase the sensitivity due to the beam splitter's change in size, resulting in an effective suppression factor in the optical path length alteration with respect to the GW strain sensitivity. Namely, the $N\gg1$ to-and-back passages across the Fabry-P\'erot cavity yield a factor of $1/N_{\text{eff}}$ such that the optical path length is given by\cite{PhysRevResearch.1.033187}
\begin{equation}\label{equ:nomirror}
    \delta(L_x-L_y) \approx \frac{1}{N_{\text{eff}}}\sqrt{2}nl\left( -\frac{\delta \alpha}{\alpha} -\frac{\delta m_\Psi}{m_\Psi}\right)\,,
\end{equation}
where $n$ is the refraction index, $l$ is the thickness of the beam splitter and $N_{\text{eff}}\sim \mathcal{O}(N)$. We emphasize here that this naturally holds only for the beam splitter. The mirrors are not affected by the suppression factor as the laser beam reflects them for every one of the $N$ to-and-back passages. The factor of $\sqrt{2}$ in Eq. \eqref{equ:nomirror} results from geometric considerations (see \cite{PhysRevResearch.1.033187} and references therein). In this format, Eq. \eqref{equ:nomirror} does not include effects on the mirrors. They will only be influenced in width and the impact on the effective optical pathway is given by the difference between the mirror changes. This is particularly relevant as the TDM field effects are highly local. Adding mirror effects to the optical path change \eqref{equ:nomirror}, the complete effect on the optical pathway yields
\begin{multline}\label{equ:-1}
    \delta(L_x-L_y) \approx \sqrt{2}\frac{g_{\textsc{DW}}}{N_{\text{eff}}} n l \left.\frac{\phi^2}{f^2}\right|_{\text{beam-splitter}} \\
    + wg_{\textsc{DW}} \left( \left.\frac{\phi^2}{f^2}\right|_{\text{y-mirror}} - \left.\frac{\phi^2}{f^2}\right|_{\text{x-mirror}} \right)\,.
\end{multline}
Here, the width of the mirrors in a $\phi$-free vacuum is denoted by $w$. Each term is evaluated at the position indicated by the subscript. Note that in Eq. \eqref{equ:-1} we disregard any effects resulting from the couplings of the DM field to fermionic mass terms. They can be easily accounted for by adding terms similar to those in \eqref{equ:-1} but replacing the coupling constants applying the prescription $g_{\text{DW}}\rightarrow\lambda_{\text{DW}}$. 
Let us point out here again that in the latter equation $L_{x,y}$ are the \emph{physical} arm lengths without the multiplication factor resulting from the Fabry-P\'erot cavities\footnote{The notation used here may exhibit variations across the literature. It is worth noting that in some instances, the suppression factor may be omitted.}.

With Eq. \eqref{equ:-1} in hand, we can now determine the resulting phase difference which is the physical quantity that is measured over time by a GW interferometer such as the LIGO detector. The phase shift $\Delta \phi$ is approximated by 
\begin{equation}
    \Delta L \approx \frac{\lambda }{2\pi}\Delta \varphi\;,
\end{equation}
where the left-hand side can be replaced by \eqref{equ:-1}. Thus, we find
\begin{multline}\label{equ:beam_mirror}
     \Delta \varphi \approx \frac{2\pi}{\lambda}\left[ \sqrt{2} n l \left.\frac{g_{\textsc{DW}}}{N_{\text{eff}}}\frac{\phi^2}{f^2}\right|_{\text{beam-splitter}} \right.\\ \left. + w g_{\textsc{DW}}\left( \left.\frac{\phi^2}{f^2}\right|_{\text{y-mirror}} - \left.\frac{\phi^2}{f^2}\right|_{\text{x-mirror}} \right)\right]\,,
\end{multline}
where the wavelength $\lambda$ is the laser wavelength of the interferometer. In this form, Eq. \eqref{equ:beam_mirror} covers all relevant effects from optical components included in LIGO, Virgo and KAGRA. For a detailed discussion, see \cite{PhysRevResearch.1.033187}. Generally, for instrumental setups as depicted in Fig. \ref{fig:RecycleFPI} an additional term contributes to \eqref{equ:beam_mirror} in case the thickness difference of 'ETMX' and 'ITMX' with respect to the mirror thickness ``ETMY" and ``ITMY", denoted by $\Delta w$, is significant. Then, one needs to add
\begin{align}
    \delta (L_x -L_y) \approx -\Delta w \left(-\frac{\delta \alpha}{\alpha} -\frac{\delta m_\Psi}{m_\Psi}\right).
\end{align}
For the instruments considered in this work, the latter is negligible. In LIGO for instance, $|\Delta w|\approx 80$ $\mu$m rendering the contribution of the latter addition irrelevant due to a suppression by multiple orders of magnitude. We emphasize at this point that increasing $\Delta w$ may be a potential way to significantly increase the interferometer sensitivity with respect to time-dependent changes in $\alpha, m_e$, constituting a concrete hardware modification fostering the detectability of TDM with interferometer technology.

For interaction terms modifying the fine-structure constant, naturally, current experimental bounds have to be satisfied by $\delta \alpha$ (see \cite{10.1063/5.0105159} for the latest test). Thus, when choosing relevant couplings caution is necessary as each DW passing can result in drastic changes of $\alpha$ if the coupling constants are not fine-tuned such that $\delta\alpha$ stays within its current measurement uncertainties. Alternatively, one can find an interaction for which the modification of the fine-structure constant vanishes far away from the DW. An interaction meeting such criteria would be, for instance, \cite{Khoze_2022}
\begin{equation}\label{equ:the_interaction}
     \mathcal{L}_{int}\supset \frac{1}{4}g_{\textsc{DW}} \sin^2 \left(\frac{N_A\phi}{f}\right) F_{\mu\nu}F^{\mu\nu}\;.
\end{equation}
Note that upon inserting Eq. \eqref{equ:2}, we have to introduce another integer quantity, $N_A$, to ensure vanishing effects far away from the DW. Doing so, the fraction $N_A/N_\phi$ has to be either integer or half-integer.

The type of interaction outlined in \eqref{equ:0} and \eqref{equ:the_interaction} introduces a temporal phase difference of the detector arms solely by modifying the size and refraction index of the beam splitter and mirrors. However, Eqs. \eqref{equ:0} and \eqref{equ:the_interaction} and other interactions also affect the propagation of the laser beams within the interferometer arms via a modified dispersion relation in case TDM is present inside the cavity. 

To demonstrate the effects resulting from a DM modification of the dispersion relation and to cover a larger part of the theory space, we introduce two additional interaction types with a coupling to a photon field in QED. Concentrating on photon operators of dimension $\leq 4$, we can build an effective photon mass term coupling via
\begin{equation}\label{equ:4}
    \mathcal{L}_{int} \supset \frac{1}{2}m_0^2 \sin^2\left(\frac{N_A \phi }{f}\right)A_\mu A^\mu\,,
\end{equation}
where $m_\gamma^2 = m_0^2 \sin^2\left(\frac{N_A \phi}{f}\right)$, 
and an axion-like\footnote{Note that the term ``axion-like'' can vary slightly in meaning depending on the reference.} coupling,
\begin{equation}\label{equ:5}
     \mathcal{L}_{int}\supset \frac{1}{4} \Tilde{g}_{\textsc{DW}} (\phi/f)F_{\mu\nu}\Tilde{F}^{\mu\nu}\,.
\end{equation}
These three interactions are most commonly used in existing literature and will modify the dispersion relation of photons when applied to the QED sector. Note here that, while \eqref{equ:5} affects the fine-structure constant similarly to \eqref{equ:0}, the interaction \eqref{equ:4} does not.

Let us now consider modifications of the dispersion relation caused by the interactions \eqref{equ:0}, \eqref{equ:4}, and \eqref{equ:5}. For demonstrative purposes, let us treat Eq. \eqref{equ:4} to provide exemplary equations. It is easy to see that
\begin{equation}
    \omega^2 = k^2  + m_0^2 \sin^2 \left(\frac{N_A\phi}{f}\right)\,,
\end{equation}
such that an altered phase velocity $v_p = \frac{\omega}{k}$ is obtained where
\begin{equation}\label{equ:6}
    v_p = 1 + \frac{m_0^2}{2k^2}\sin^2 \left(\frac{N_A\phi}{f}\right)\;.
\end{equation}
It is easy to see that Eqs. \eqref{equ:0} and \eqref{equ:5} yield similar modifications, namely
\begin{subequations}
\begin{equation}
    \omega^2 - k^2 = \pm \frac{\Tilde{g}_{\textsc{DW}}}{f}(\omega \partial_t \phi + \Vec{k}\Vec{\nabla}\phi)\,,
    \end{equation}
    \begin{equation}
    \omega^2 - k^2 = i \frac{g_{\textsc{DW}}}{f}(\omega \partial_t \phi^2 + \Vec{k}\Vec{\nabla}\phi^2)\,.
\end{equation}
\end{subequations}
In the case where $A_\mu$ is the $U(1)$ gauge potential, the imaginary factor in the latter equation leads to an exponential power in the plane electromagnetic wave. This undesirable feature can be suppressed by choosing a sufficiently small $g_{\textsc{DW}}$. However, this effect sourced by $g_{\rm DW}$ or $\Tilde{g}_{\rm DW}$ is subdominant to the phase shift caused by the size changes for the ground-based GW detectors~\cite{Khoze_2022}. Therefore we only consider dispersion effects sourced by $m_0$.

Starting from Eq. \eqref{equ:6}, both sides of the equation can be integrated such that we obtain
\begin{equation}
    2L = t- t_0 + \frac{m_0^2}{2\omega^2}\int_{t_0}^t \dd t \sin^2 \left(\frac{N_A\phi}{f}\right)\;,
\end{equation}
where we set $c=1$. Here we replaced $k^2$ with $\omega^2$ with respect to \eqref{equ:6} as we expand in zeroth order in $m_0$ in the denominator. Note that this formula holds for both arms equivalently. The dynamic information of the DW is solely encoded in the TDM solution $\phi$. Hence, we replace $\phi$ with $\phi\equiv \phi_{\text{TDM}}(\mathbf{x})$, where $\mathbf{x}$ contains information about the trajectory and $\phi_{\textsc{TDM}}(\mathbf{x})$ describes the extension of the DW. Outside of the DW, the integrand vanishes.

Finally, we consider the signal due to a displacement of the center of mass induced in the test masses due to spatial gradients in the fundamental constants \cite{PhysRevResearch.1.033187}. The latter gives rise to an acceleration of the test masses with mass $m_{\rm test}$ given by
\begin{align}
\label{equ:delta_a}
    \delta a = - \frac{\nabla m_{\rm test}}{m_{\rm test}}\,.
\end{align}
Overall, this acceleration is induced by three distinct effects: the mass change of the nucleus, the mass change of the electron, and the change in electrostatic repulsion between protons \cite{PhysRevResearch.1.033187}, which can be induced by a change in $\alpha$. Individually, these three contributions differ only in amplitude, not in signal shape. We can consider an interaction like \eqref{equ:0}; but with vanishing changes far from the DW as
\begin{multline}\label{equ:the_interaction2}
\mathcal{L}_{int}\supset \frac{1}{4}g_{\rm DW}\sin^2\left(\frac{N_A \phi }{f}\right)F_{\mu\nu}F^{\mu\nu}\\+\sum_i\lambda_{i} \sin^2\left(\frac{N_A \phi }{f}\right) m_i \Psi_i \bar \Psi_i.
\end{multline}
For ground-based detectors, this effect is dominant over the size changes. Size changes are dominant over the dispersion effects induced by the same term as well. Consequently, we consider the center-of-mass change as the only effect induced by $g_{\rm DW}$ and $\lambda_i$. The magnitude of the center-of-mass displacements from each of the three contributions are found by scaling their individual effects by their fractional contribution to the total mass of the mirrors which are about 1, $3\times10^{-4}$, $1.4\times10^{-3}$ for nucleon masses, electron mass, and electrostatic energy, respectively~\cite{PhysRevResearch.1.033187}. Because the difference between them is only a scaling, the three effects are not separable. Hence, in the rest of the article we consider a single interaction caused by $g_{\rm DW}$ that effectively affects the whole mass. That constant can be converted to couplings of each constituent by scaling with their corresponding mass fractions and fermion masses.

Depending on its thickness and angle of impact with respect to the plane defined by the interferometer arms, the signatures of the resulting phase shift in the $x$ and $y$ directions can be very distinct (we adopt the same labeling as in Fig. \ref{fig:RecycleFPI}). Further, we want to take into account a complete signal, i.e., we want to obtain a numerical solution for the time interval starting from $t_0$, where the DW does not intersect with the detector, to $t_1$, where the DW has passed through it.

\subsection{Geometrical considerations}
\begin{figure}
    \centering
    \includegraphics[width=0.9\columnwidth]{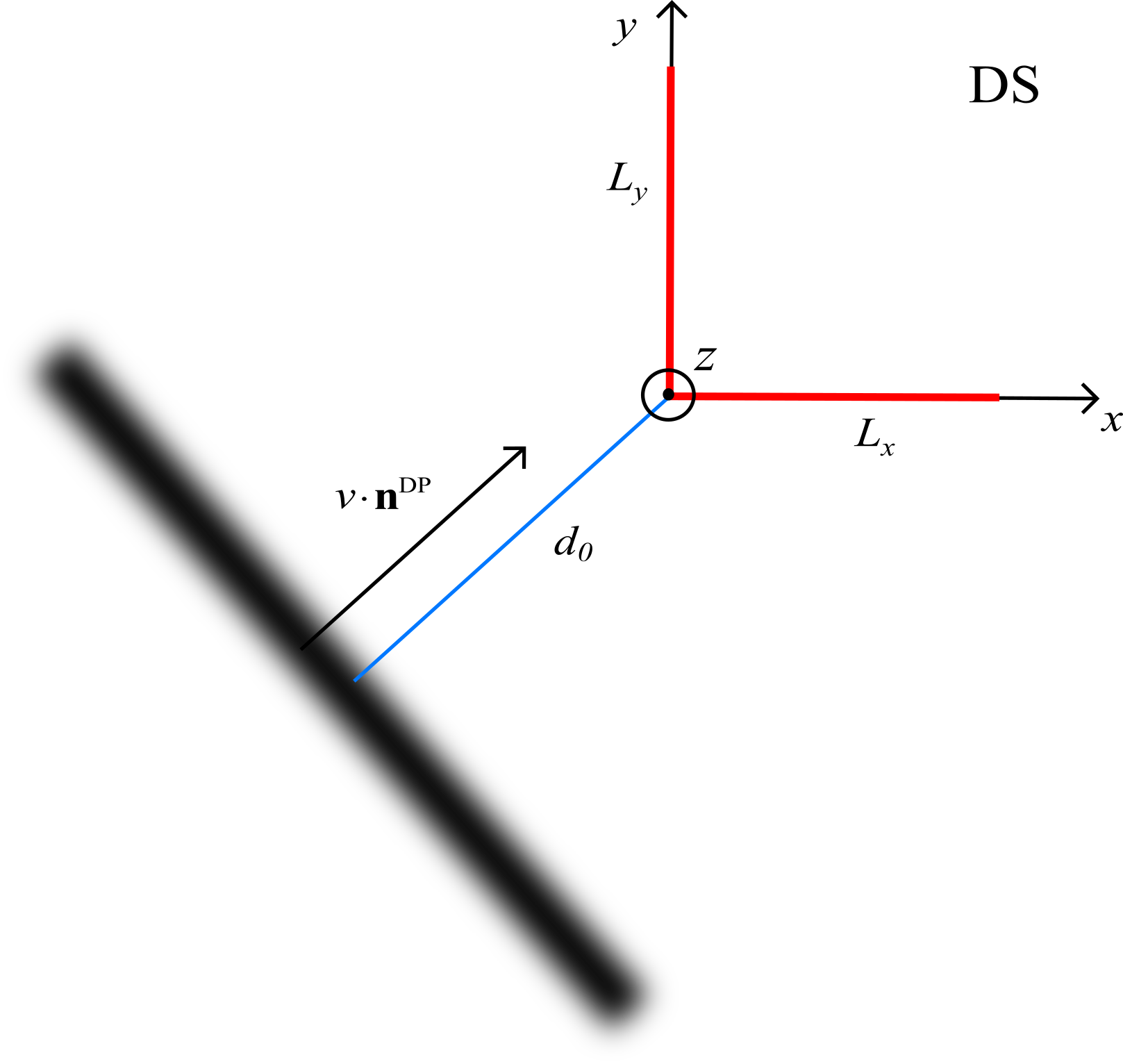}
    \caption{Two-dimensional sketch of a DW passing through a GW interferometer. The plane resembles the detector plane with its arms aligned with the $x,y$ axes. The initial distance and velocity are given by $d_0$ and $v$ respectively. The vector $\mathbf{n}^{\text{DP}}$ represents the projection of the DW normal onto the detector plane.}
    \label{fig:sketch}
\end{figure}

For the general analysis, we assume that the spatial extent of the DW is determined by the thickness $d$, which in turn is governed by the mass as $d\sim 2/m_{\text{dark}}$. We set the coordinate origin in our consideration to align with the center of our detector. For the remainder of this article, we label this coordinate system the detector system ($DS$). Hence, the interferometer arms align with the $x$ and $y$ axis, respectively. We define a vector $\mathbf{n}$ as the unit normal vector of the DW, which characterizes the direction of motion. Typically, one assumes a velocity around $v\sim 3 \cdot 10^{-3}c$. In our analysis, we treat the speed as well as all other model-specific parameters as free. To determine the shape, duration, and frequency of the signal produced by a modified dispersion relation, we can project the normal vector $\mathbf{n}$ onto the detector plane such that the direct distance, i.e.,
\begin{equation}\label{equ:8}
    \Tilde{\mathbf{x}}(t) := \mathbf{x}\cdot\mathbf{n}^\textsc{DP} - d_0 + vt\;,
\end{equation}
can be inserted into the solution \eqref{equ:2}.
Here, $d_0$ denotes the initial distance between the DW and the detector measured normal to the plane. In turn, $\mathbf{n}_{x,y}^\textsc{DP}$ is the projection of the normal vector onto the detector plane, which can be parametrized by two angles such that $\mathbf{n}_{\textsc{DP}}= (\sin \theta \cos \phi, \sin \theta \sin \phi )$ (In Fig. \ref{fig:sketch}, we schematically sketch the complete DM-detector configuration.)- The angles are given in the $DS$. The vector $\mathbf{x}$ labels an arbitrary point in space. In this setup, Eq. \eqref{equ:8} labels the distance between the DW and an arbitrary point $\mathbf{x}$ at time $t$. To obtain a dynamical description of the DW, Eq. \eqref{equ:8} can be directly inserted into \eqref{equ:2}. Note that in doing so, we automatically adopt the assumption of infinitely extending planar dimensions of the DW. Given that the large spatial dimensions of a topological defect are set by the length scale of causally connected regions in spacetime at the symmetry-breaking phase transition, the latter assumption is well justified.

Given the geometrical setup in Fig. \ref{fig:sketch}, consider a modification of the phase velocity in terms of some function of the dark scalar $\phi_{\text{TDM}}$ and a photon wave vector $\mathbf{k}$, i.e.,
\begin{equation}
    v_p = \frac{\omega}{k} = 1 + \gamma(\phi_{\text{TDM}}[\Tilde{\mathbf{x}}(t)],\mathbf{k}) \approx 1 + \gamma(\phi_{\text{TDM}}[\Tilde{\mathbf{x}}(t)],\mathbf{\omega})
\end{equation}
where in the last step we replaced $\mathbf{k}$ with $\omega$ which, again, holds to zeroth order in $m_0\ll1$.
Integrating both sides of the latter equation, one finds
\begin{equation}\label{equ:6.5}
    N_{\text{eff}}L = t-t_0 + \int_{t_0}^{t} \dd t' \gamma(\phi_{\text{TDM}}[\Tilde{\mathbf{x}}(t')],\mathbf{\omega})\;,
\end{equation}
where $t_0$ is the time when a photon emitted from the laser leaves the beam splitter for the first time and $t$ is the time when it returns from the Fabry-P\'erot cavity and meets the beam splitter again. Depending on the construction of the instrument, it will pass through the cavity with arm lengths $L$ roughly $N_{\text{eff}}$ times. Hence, we gain a factor on the left-hand side. Note that, by similar arguments as before, here we find that the signal induced by $\phi_{\text{TDM}}$ is multiplied by the number of to-and-back passages inside the cavities instead of being suppressed as for the size change of the beam splitter for instruments with hardware configurations similar to Fig. \ref{fig:sketch}. Based on a plane-wave representation of the photons inside the cavity, we can then use \eqref{equ:6.5} to compute the phase difference in each arm as
\begin{equation}\label{equ:7}
    \Delta \varphi_{x_i}(t) = \omega \int_{t-N_{\textsc{eff}}L}^{t} \dd t' \gamma(\phi_{\text{TDM}}[\Tilde{x}_{x_i}(t')],\mathbf{\omega})\;.
\end{equation}
Note that, since we aligned our coordinate system with the detector arms we can split $\Tilde{\mathbf{x}}$ into its $x$ and $y$ components and completely separate Eq. \eqref{equ:6.5}. This results from the chosen alignment: in the reference frame $DS$, the DW travels with the velocity $v\ll c$ in the direction $\mathbf{n}$. Projected onto the detector plane, only the $x,y$ components of the normal vector of the DW are relevant, i.e., $\mathbf{n}_{\textsc{DP}}= (\sin \theta \cos \phi, \sin \theta \sin \phi )$. For an arbitrary point in space, $\mathbf{x}$, the field strength of the TDM is given by $\phi_{\text{TDM}}(\mathbf{x}\cdot\mathbf{n}_\textsc{DP} - d_0 + vt)$. In our consideration, the only relevant points in spacetime are the ones residing along the detector arms which are by definition of $DS$ aligned with the coordinate axis. Thus, calculating $\Delta \varphi$ along the $x$ axis, the $y$ component of the scalar product $\mathbf{x}\cdot \mathbf{n}^{\text{DP}}$ vanishes, and so $\Tilde{x}_{x_i}(t') = x_i n_{i}^{\text{DP}}- d_0 + vt$ for $x_i\in\{x,y\}$.

Finally, to compute the physical quantity appearing in actual measurement data from cutting-edge interferometers, we compute the phase difference between the interferometer arms,
\begin{equation}
    \Delta \varphi(t) = \Delta \varphi_x(t)-\Delta \varphi_y(t)\,,
\end{equation}
and find 
\begin{multline}
    \Delta \varphi (t) = \omega N_{\text{eff}} \left(\int_{L_x} \dd x \,\gamma(\phi_{\text{TDM}}[x \sin\theta \cos \phi - d_0 - vt ]) \right. \\ \left. -  \int_{L_y} \dd y \,\gamma(\phi_{\text{TDM}}[y \sin\theta \sin \phi - d_0 - vt ]) \right)
\end{multline} 
As an example, consider the interaction \eqref{equ:4} where the corresponding $\gamma(\phi) = \frac{m_0^2}{2\omega^2}\sin^2 \left(\frac{N_A\phi}{f}\right)$. Upon inserting $\phi_{\textsc{TDW}}$ one finds that the phase difference measured by the interferometer is
\begin{multline}\label{eq:phase_vel}
    \Delta \varphi (t) =  \frac{m_0^2N_{\text{eff}}}{2\omega}\\ \times \left[\int_{L_x} \dd x \,\sin^2\left( 4\frac{N_A}{N_\phi} \arctan(e^{m(x \sin\theta \cos \phi - d_0 - vt)}) \right)\right. \\ -\left. \int_{L_y} \dd y \,\sin^2\left( 4\frac{N_A}{N_\phi} \arctan(e^{m(y \sin\theta \sin \phi - d_0 - vt)}) \right)\right]\;.
\end{multline}    
The latter equation encapsulates the complete dynamical effect due to a modified dispersion relation caused by TDM interacting with the Standard Model via the interaction \eqref{equ:4}.\footnote{To repack the information contained in the latter equation into a more experimentally relevant language, it can be converted into frequency space by simply applying a straightforward Fourier transform.}

To make our analysis more robust, we consider a second detector being traversed by the TDM and located at a distance $D$ with respect to the first detector. Together both detectors resemble the LIGO interferometers located in Livingston and Hanford. A DW of ``infinite'' planar spatial extend will inevitably introduce a phase shift $\Delta \varphi$ in both detectors. In principle, the orientation and distance of the two interferometers determines the correlation time and the shapes of the signals exactly, based on the knowledge of $\mathbf{n}$ and the DW's velocity. 

\begin{figure}
    \centering
    \includegraphics[width=0.9\columnwidth]{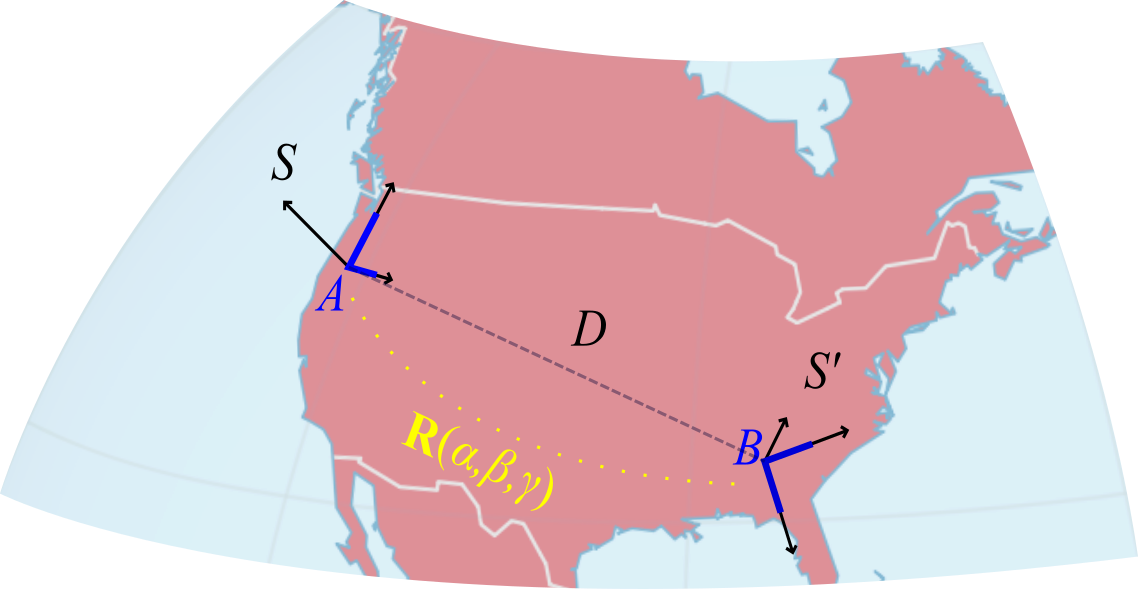}
    \caption{Sketch of the experimental setup outlined in this section. The two detectors $A,B$ are chosen to resemble the LIGO detectors at Hanford and Livingston. Note that this sketch does not respect scaling or alignment.}
    \label{fig:sketch2}
\end{figure}

To consider a two-detector setup, we lay out our coordinate system $S$ by defining a detector $A$, which we place at the origin of the coordinate grid such that the arms are aligned with the $x,y$ axes. In this frame, let $D$ be the direct distance to another coordinate system $S'$ at whose origin the second detector $B$ is placed, such that the arms again are aligned with the $x,y$ axes. Also, let $\mathbf{d}$ denote the vector connecting the origins, i.e., $\mathbf{d} = \mathbf{0}'-\mathbf{0}$. Due to the Earth's curvature, we find the two systems to be not related by translations but also rotated with respect to each other, where we define $\mathbf{R}$ to be the rotation matrix. While the signal in detector $A$ is determined by the projection of $\mathbf{n}$ onto the detector plane, i.e., the $x,y$ plane in $S$, the signal in $B$ is determined by the projection of $\mathbf{R}\mathbf{n}$ onto the $x,y$ plane in $S'$. The time in between the signals appearing in $A$ and then in $B$ or vice versa depends on the orientation of $\mathbf{n}$ with respect to the translation $\mathbf{d}$ connecting $S$ and $S'$. If they are $\mathbf{d}\perp \mathbf{n}$, the signals appear in both detectors simultaneously. On the other hand, if $\mathbf{d}\parallel \mathbf{n}$, the signals are correlated over a $D/v$ time interval. In between these two thresholds, we expect a nonuniform random distribution of correlation times $t_\text{corr} \in [0, D/v]$. Figure \ref{fig:sketch2} illustrates the positions of the two detectors.

Let us now summarize the configuration described in this section which we use throughout the following analysis in Section \ref{ana}. We consider TDM in the form of a DW \eqref{equ:2} passing through the Earth. Its direction and speed are given by $\mathbf{n}$ and $v$, respectively. For simplicity, the spatial extent of the DW perpendicular to $\mathbf{n}$ is assumed to be infinite. By passing through the Earth, the DW leaves a trace in two detectors being placed on its surface with a direct distance $D = |\mathbf{d}|$. Since the detectors' arms are not aligned, the phase shifts induced by the passing DW will be of different shape and frequency, and the signals measured by the individual instruments will be delayed with respect to each other. In general, the TDM traversing a detector yields two types of effects: a modification of the dispersion relation and a change of the optical pathway due to the modified spatial extent of the beam splitter and the mirrors of the interferometer. The exact shape of the phase shift induced depends on the type of interaction and hence on the underlying theory.

\subsection{Topological dark matter signals in detectors}
\label{signals_TDM}

\begin{figure}
   \includegraphics[width=0.95\columnwidth]{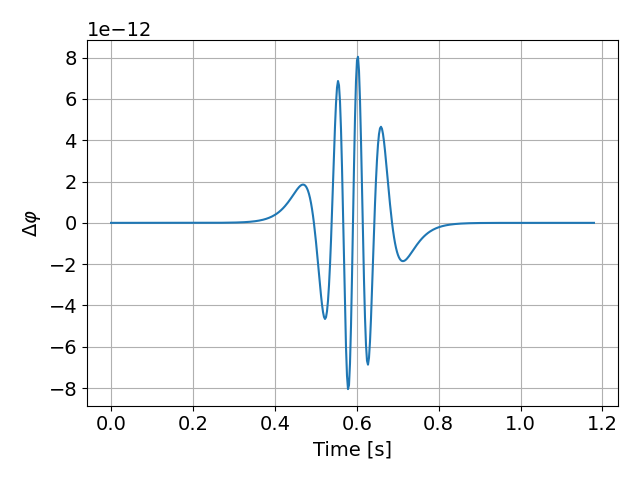}    

\includegraphics[width=0.95\columnwidth]{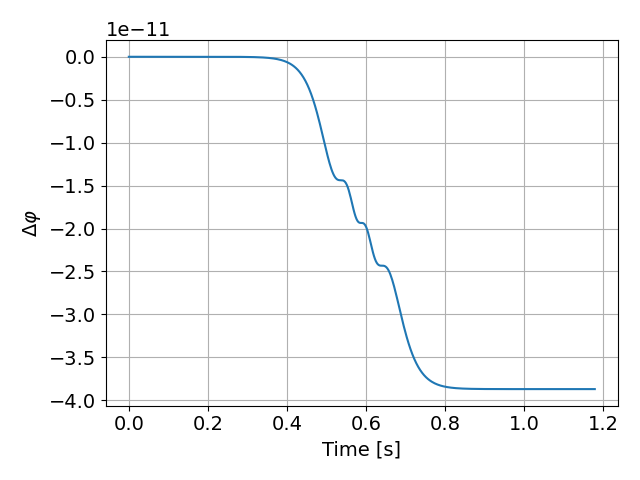}   
     \caption{Exemplary phase difference signals (top) from dispersion from Eq. \eqref{equ:4} and (bottom) from center-of-mass changes from \eqref{equ:the_interaction2}, considering the impact on the change in electron masses only (bottom). Here, we consider the LIGO Hanford detector with $m_\text{dark}= 10^{-12}$ eV, $m_0=10^{-10}$ eV, $N_A/N_\phi=2$, $N_{\text{eff}}\approx 300$, $v = 7\cdot 10^{-2}$, and $g_{\rm DW}=10^{-25}$.}
       \label{fig:sig_examples}
\end{figure}

The discussed interactions boil down to two distinct classes of signals that can be produced by TDM, displayed in Fig. \ref{fig:sig_examples}.
In Fig. \ref{fig:sig_examples}, the upper plot shows a signal due to the modification of the dispersion relation via $\mathcal{L}_{\rm int}\sim m_0^2\sin^2{(\chi\phi)A_\mu A^\mu}$. The amplitude of the phase shift (and, correspondingly, the dominance of interactions) depends strongly on the selected coupling constants $m_0$. The number of oscillations and asymmetry depends on the fraction $N_A/N_\phi$ and the angle by which the DM surpasses the detector plane respectively. The bottom plot in Fig. \ref{fig:sig_examples} displays the generated signal from center-of-mass displacements. Strikingly, it exhibits similarities to the nonoscillatory (or electric) part of the GW memory effect \cite{Nonlin_mem}. The latter is difficult to detect in the considered instruments due to their natural sensitivity cutoff for lower frequencies \cite{Grant_2023}. Nonetheless, signals of this type cannot be excluded in our investigation \textit{a priori}. Thus, in Sec. \ref{ana} we include them in our parameter estimation scheme. Overall, we have two types of dominant signals: dispersion-induced signals from \eqref{equ:4} and center-of-mass changes induced from \eqref{equ:the_interaction2}. All parameters relevant to these interactions are bounded by observational constraints (if available) but otherwise treated as free fit parameters.

\begin{figure}
\includegraphics[width=0.99\columnwidth]{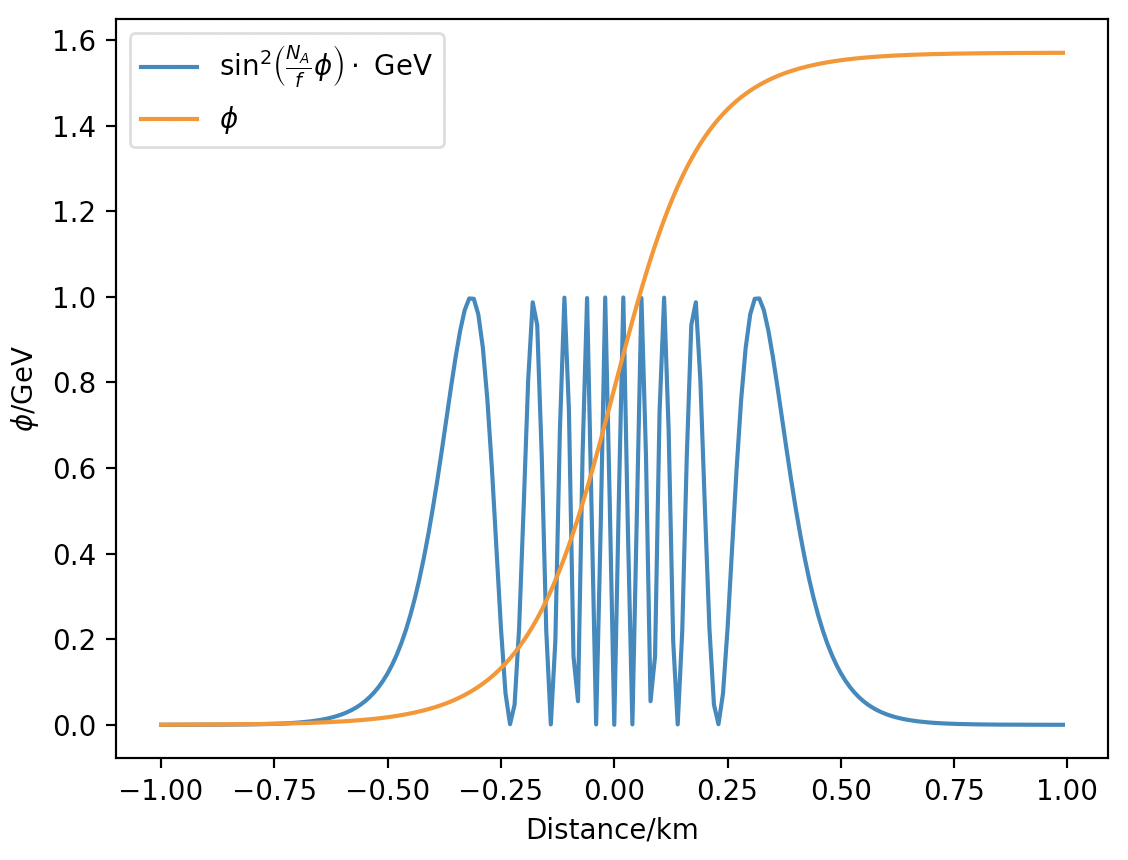}   
     \caption{Exemplary configuration for the TDM solution \eqref{equ:2} with $m_\text{dark}=10$ neV, $f=1$ GeV, and $N_\phi =4$.}
       \label{fig:Phi_examples}
\end{figure}

To provide a more intuitive explanation for the signals' morphology displayed in Fig. \ref{fig:sig_examples}, we show an exemplary field configuration for TDM [and $\sin^2(\chi \phi_{\text{TDM}})$] in Fig. \ref{fig:Phi_examples}. The combination of $\sin^2(\chi \phi_{\text{TDM}})$ in the interaction term \eqref{equ:4} together with the $\arctan$ in the stable DW solution for DM, $\phi_\text{TDM}$, leads to a symmetric oscillation (blue line in Fig. \ref{fig:Phi_examples}). The number of maxima depends mainly on $N_A/N_\phi$. At the peaks of $\sin^2(\chi \phi_{\text{TDM}})$ the dispersion relation is modified the strongest [see \eqref{equ:6}]. Virtually sending this configuration through the interferometer, there will generally be an unequal distribution of peaks intersecting with $L_x$ compared to $L_y$, causing different modifications in the dispersion $\omega^2$. This causes the oscillatory phase behavior depicted in the top plot of Fig. \ref{fig:sig_examples}.

\subsection{Sensitivity estimation}
\begin{figure}
    \centering
    \includegraphics[width=\columnwidth]{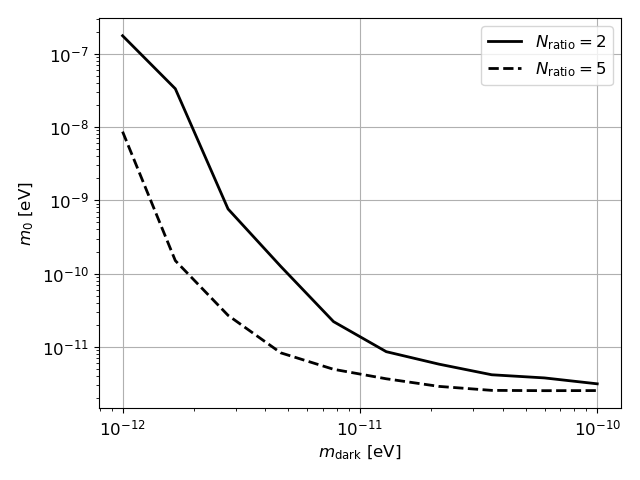}
    \includegraphics[width=\columnwidth]{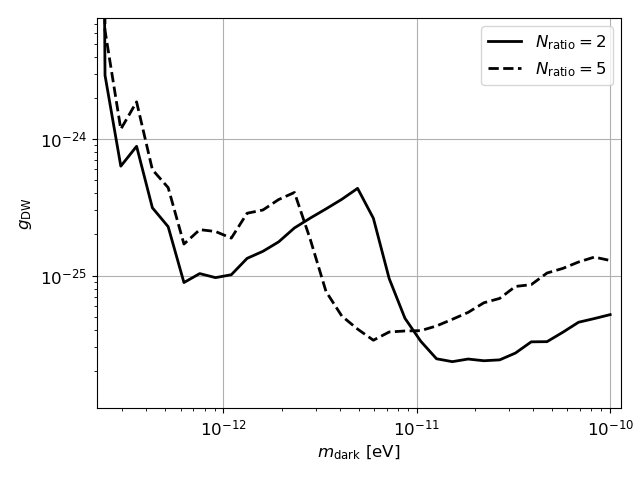}
    \caption{90\% sensitivities of the couplings as a function of the scalar mass and the $N_{\rm ratio}$ parameter for $v_{\rm DW}=0.01c$}
    \label{fig:upper}
\end{figure}

Prior to the data analysis, we estimate the potential sensitivity of the search. To do so, we compute the so-called 90\% sensitivities for the couplings $m_0$ and $g_{\rm DW}$, representing values of these coupling constants for which a signal would be more significant than the median of the background 90\% of the time. Alternatively, these values can be interpreted as 90\% upper limits that can be established following a search resulting in a $p$-value of 0.5. We calculate the sensitivities as a function of the scalar mass $m_{\rm dark}$ for two example cases of $N_{\rm ratio}=N_A/N_\phi=\{2,5\}$, assuming a DW speed of $0.01c$ and considering the couplings separately.  The detection statistic for significance is assumed to be the network signal-to-noise ratio (SNR) from two LIGO detectors. For each scalar mass and $N_{\rm ratio}$,  we conduct a matched-filtering search for those specific parameters while exploring 400 uniformly distributed and equally spaced directions in the sky. Gaussian noise at the sensitivity level of LIGO detectors during the O3 run is considered for the background distribution, which consists of the results of the matched filtering search in the absence of a real signal. Although the presence of glitches in the data may affect sensitivity, the median of a population is generally robust against outliers, minimizing their impact. The obtained sensitivities are shown in Fig. \ref{fig:upper}. The boundaries of the scalar mass are determined by the sensitive frequency band of LIGO and the sampling rate. Lower masses could result in signals with frequencies falling outside of the LIGO band due to the rapidly increasing seismic noise below $10$~Hz, while higher masses would yield high-frequency signals necessitating sampling rates higher than $\gtrsim10$~kHz. Overall, the sensitivities for the coupling constants crudely decrease with increasing $m_{\rm dark}$, as expected. Their exact shapes depend on the sensitivity of the detectors to different frequencies and the signal shapes generated by the interactions. Sensitivities for different DW speeds  ($v_{\rm DW}$) can be derived by scaling the parameters that satisfy the given equalities

\begin{subequations}
    \begin{equation}
        m_{\rm dark}v_{\rm DW}={\rm constant}_1\,,
    \end{equation}
    \begin{equation}
        m_{\rm dark}m_0^2={\rm constant}_2\ {\rm or}\ m_{\rm dark}g_{\rm DW}={\rm constant}_2\;.
    \end{equation}
\end{subequations}
Parameters satisfying the above equations would produce the same signal. We find that the sensitivities range from $\sim 10^{-7}-10^{-12}$ eV for $m_0$ and from $\sim 10^{-25}-10^{-24}$ for $g_{\rm DW}$ in the $m_{\rm dark}$ mass range $10^{-12}-10^{-10}$ eV for $v_{\rm DW}=0.01c$.    

This sensitivity estimation can be utilized as a gauge for the subsequent analysis in Sec. \ref{ana}. There, the interaction in \eqref{equ:4} and the resulting effects on hardware and laser propagation are analyzed numerically and compared to massive BBH events and glitches. We stick closely to the configuration outlined above where the two detectors in Fig. \ref{fig:sketch2} are assumed to be the LIGO detectors in Hanford and Livingston. Note that we do not include a third detector, such as Virgo or KAGRA, in our investigation as we found that for most applied cases the SNR is too low to carry relevance for potential detections of TDM, i.e., it will not affect the fitting procedure. 

%%%%%%%%%%%%%%%%%%%%%%%%%%%%%%%%%%%%%%%%%%%%%%%%%%%%%%%%%%%%%%%%%%%
                    %%
                    %%
%%%%%%%%%%%%%%%%%%%%%%%%%%%%%%%%%%%%%%%%%%%%%%%%%%%%%%%%%%%%%%%%%%%
\section{Analysis}\label{ana}
Employing numerical simulations of the phase shifts resulting from the aforementioned interactions, we examine the potential existence of TDM within actual GW detection data. Our analysis unfolds in two phases: First, we inquire whether certain detected BBH mergers could also exhibit characteristics akin to TDM signatures as illustrated in Fig. \ref{fig:sig_examples}. Our focus is particularly directed toward events characterized by high estimated detector-frame black hole masses, given their brief signal durations within the sensitive frequency band of the detectors. These types of events are particularly interesting as they do not distinctly exhibit the entire inspiral-merger-ringdown waveform characteristic of a merging binary which would be impossible to fit with TDM signatures. We explore whether these short observed signals, which can be attributed to the final stages of a massive BBH compact binary coalescence (CBC), may also be accounted for by the passage of TDM through the detector. It is important to acknowledge that this analysis inherently carries a negative selection bias towards testing the TDM scenario, as these events were identified through template-based searches employing BBH waveforms. Consequently, they are anticipated to resemble BBH waveforms \textit{a priori}. Nonetheless, we undertake this study as the inaugural analysis assessing the presence of TDM signals in the data. A comprehensive search for such signals would necessitate a specifically tailored pipeline, as we elaborate in Sec. \ref{discus}.

In the second part of this analysis, we explore the resemblance between TDM signals and noise artifacts in the data, known as glitches. This second phase serves as an exploratory analysis for a TDM search within the GW data, as TDM waveforms resembling glitches could potentially be easily discarded, complicating their identification. We choose a particular set of glitches to test TDM signals against. Our selection criteria rely on assessing the degree of similarity between TDM signals and glitches in frequency-time diagrams.

\subsection{TDM as BBH events}\label{bbhs}
There are two primary types of searches that can potentially detect BBH mergers: template-based searches and unmodeled burst searches. Template-based searches operate under the assumption of a CBC scenario, employing matched-filtering techniques with GWs specific to CBCs. Conversely, in general, burst searches do not presuppose a particular astrophysical scenario; instead, they seek coherent excess power across different GW detectors. Over the course of the three observing runs of the Advanced LIGO and Advanced Virgo detectors, approximately 90 confident mergers were detected. All of these detections were made through template-based pipelines, implying their interpretation as CBC mergers. 

While the signals from these detections could be accounted for by CBC waveforms, the parameters of certain events were astrophysically unexpected, challenging our current understanding of stellar evolution models. One of the first and most drastic examples of such an event was GW190521~\cite{PhysRevLett.125.101102}. Being a confident detection found by both the CBC and burst pipelines, the masses of both of its component BHs were estimated to be most likely above 50 M$_{\odot}$, lying in the so-called upper mass gap or the pair-instability mass gap where no compact object is expected from stellar evolution due to the pair-instability supernova phenomenon. Alternative explanations for its origin or properties have been proposed, such as hierarchical mergers~\cite{Veske_2021,Kimball_2021,Tagawa_2021}, an eccentric merger~\cite{Romero_Shaw_2020,gamba2023gw190521,gayathri2022eccentricity}, or components with exotic matter~\cite{PhysRevLett.126.081101,bustillo2022searching}. One scenario dismissed in the original analysis was the possibility of a GW signal originating from cosmic strings rather than a BBH merger~\cite{Abbott_2020}.

The broad range of explanations for this event is a direct result of the short observed signal duration. Being a heavy-mass BBH merger, only the last few cycles of the merger were observable in the sensitive frequency range of LIGO and Virgo. As mentioned above, this property promotes GW190521 to be a prime candidate for fitting TDM signals to detector data. The similarity of events akin to GW190521 and TDM signatures is expected to be particularly strong when a chirp-like evolution is not clearly resolved in the detection data. In this case, the chirps' nonsymmetric time evolution (which is in stark contrast to the TDM signal's time-evolution symmetry) is overshadowed by noise. 

We emphasize that our analysis aiming at alternative explanations for selected events in recorded data is substantially different and structurally much simpler than the slightly modified merger scenarios such as an eccentric merger or the merger of Proca stars. We use eight distinct fit parameters ($\boldsymbol{\theta}$) in our model, including the three-dimensional velocity vector of the TDM, a reference time for the signal (equivalent to the distance to the TDM at a reference time), and the four parameters that enter the interaction Lagrangian that we introduced earlier, namely, $m_{\rm dark}$, $m_0$, $N_{\rm ratio}$, and $g_{\rm DW}$. We consider the signals produced jointly by $m_0$ and $g_{\rm DW}$ by superposing their individually produced signals. This allows us to test the complete set of signal forms that is generated by them. Furthermore, if both of the interactions are present in reality, their separate analyses would yield inaccurate results due to incomplete modeling. We paramterize the three-dimensional velocity vector with its speed magnitude, cosine of its zenith angle with respect to the Hanford detector, and its azimuth angle with respect to the $X$ arm of the Hanford detector. The CBC signals, instead, are parametrized by 15 parameters, resulting in a significantly larger parameter space. The CBC fit parameters include the total detector frame mass of BHs, their mass ratio, their dimensionless three-dimensional spin vectors with respect to the orbital momentum, the cosine of the inclination of the orbital plane with the line of sight, the coalescence phase,the cosine of declination, the right ascension, the luminosity distance to the source, the detection time and the polarization angle of the GW vector.  

For the concrete analysis, we have chosen four confident LIGO detections with high estimated masses and noise-affected chirp as expected from CBCs: GW190426\_190642, GW190521, GW191109\_010717, and GW200220\_061928. We further analyze the marginal intermediate-mass black hole triggers 200114\_020818 and 200214\_224526. We use the $32$-second long data segments from LIGO Hanford and LIGO Livingston detectors around the events sampled at $4096$ Hz from the Gravitational Wave Open Science Center (\url{https://gwosc.org}) \cite{Abbott_2023}. 
We have opt to incorporate data from Virgo due to its comparatively lower sensitivity, a decision that does not fundamentally alter our results. To mitigate the computational expenses associated with events exhibiting low-frequency ``signals'', we downsample the data to $409.6$ Hz after filtering using an ideal low-pass filter at half the sampling frequency. We whiten the data by obtaining the noise power spectral density via Welch's average periodogram method with segment lengths of $512$, using the \texttt{matplotlib} python package's \texttt{matplotlib.pyplot.psd} function~\cite{2007CSE.....9...90H}. We perform parameter estimations for the TDM scenario using our models outlined in Sec. \ref{dm_model} and the BBH scenario using the \texttt{NRSur7dq4} waveforms~\cite{Varma_2019} via the \texttt{gwsurrogate} package~\cite{2014PhRvX...4c1006F}. For sampling, we used the \texttt{dynesty} sampler~\cite{Speagle_2020} under the \texttt{Bilby} framework~\cite{2019ApJS..241...27A,2020MNRAS.499.3295R}. We note that our sampling is potentially not optimal for finding globally optimized fits and may be stuck at the local likelihood maxima of the parameter space. 

For comparing hypotheses ($H$), we rely on the Bayesian evidence ($\mathcal{B}$) or the marginal likelihood
\begin{equation}
    \mathcal{B}(H)=\int \mathcal{L}(\boldsymbol{\theta}|H)P(\boldsymbol{\theta}|H) d\boldsymbol{\theta}\;,
\end{equation}
where $P(\boldsymbol{\theta})$ is the prior distribution of the parameters and the likelihood $\mathcal{L}$ is calculated with the residual signal after the model ($h$) is subtracted from the data ($d$) by assuming Gaussian noise. We use uniform prior distributions for our fit parameters. These are parametrized appropriately such that they correspond to isotropic orientations and distributions in space. The only exception is given by the distance prior, which we take to be proportional to the distance squared. For whitened data, up to an additional constant, the logarithm of the likelihood is given by
\begin{equation}
    \ln \mathcal{L}=-\frac{1}{2}\langle d-h(\boldsymbol{\theta})|d-h(\boldsymbol{\theta})\rangle\;,
\end{equation}
where the brackets denote the inner product of vectors. The joint likelihood from two detectors is simply the product of individual likelihoods.

Testing for optimal fits of TDM models to the aforementioned event using the interactions \eqref{equ:4} and \eqref{equ:the_interaction2} as the source of TDM signal, we find that for all of the CBC events the BBH scenario is favored over the TDM scenario. The best TDM fits are produced for event GW190521. We consider the signal around 200214\_224526 in the next subsection as an example of a glitch trigger.

\begin{figure}
    \centering
    \includegraphics[width=\columnwidth]{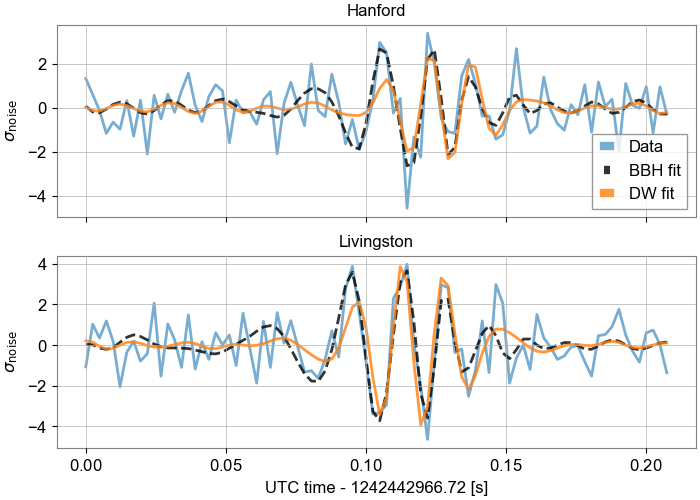}
    \caption{Whitened best-fitting TDM and BBH signals on the data of GW190521. The y axis is in units of the standard deviation of the noise.}
    \label{fig:190521}
\end{figure}

\begin{figure*}
    \centering
    \includegraphics[width=\textwidth]{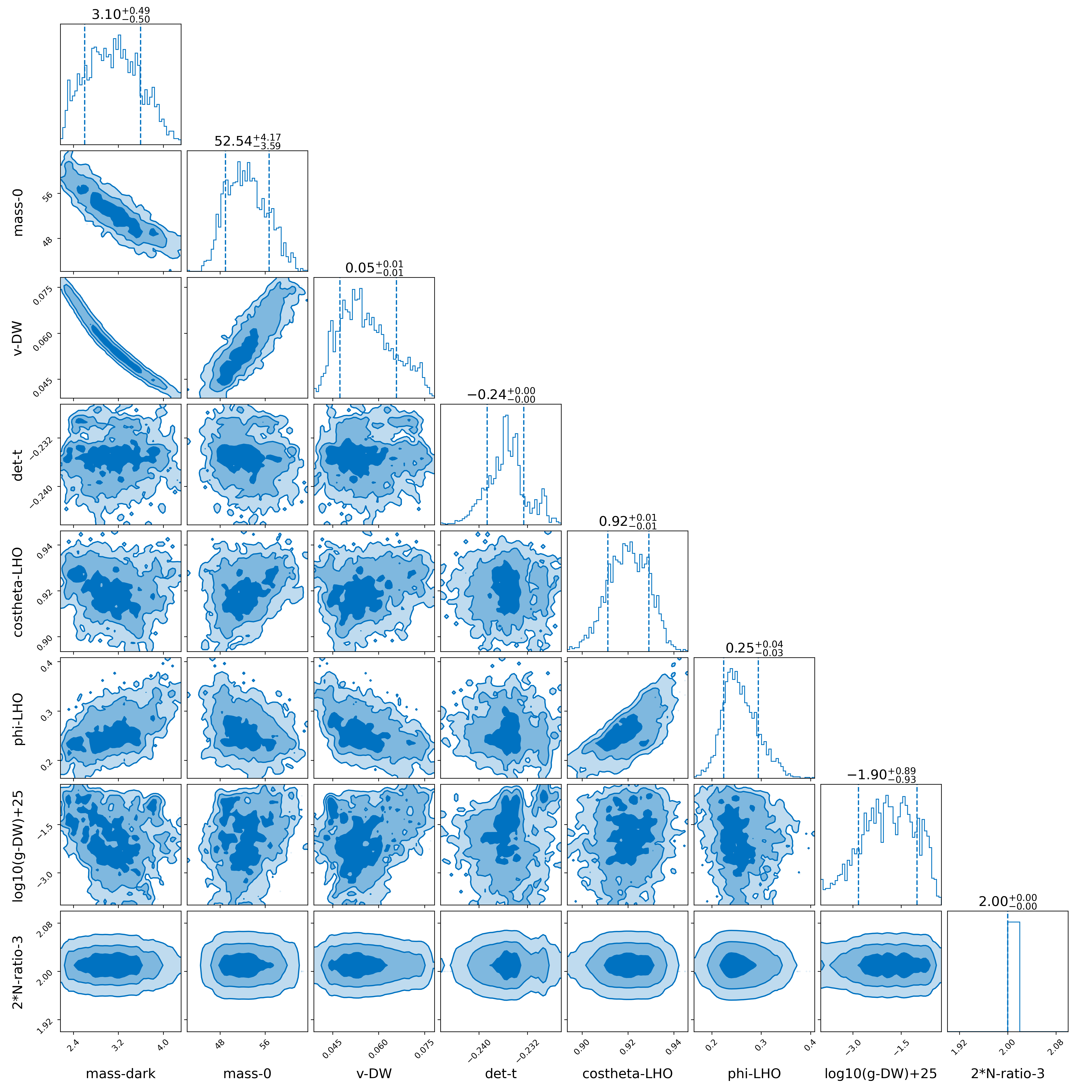}
    \caption{Parameter estimation of the TDM scenario for the GW190521. $m_{\rm dark}$ and $m_0$ are in units of $10^{-12}$ eV, $v_{\rm DW}$ in units of $c$, det-t in seconds, and phi-LHO in radians. The estimated $N_{\rm ratio}$ is 2.5.}
    \label{fig:190521PE}
\end{figure*}

\begin{figure*}
    \centering
    \includegraphics[width=\textwidth]{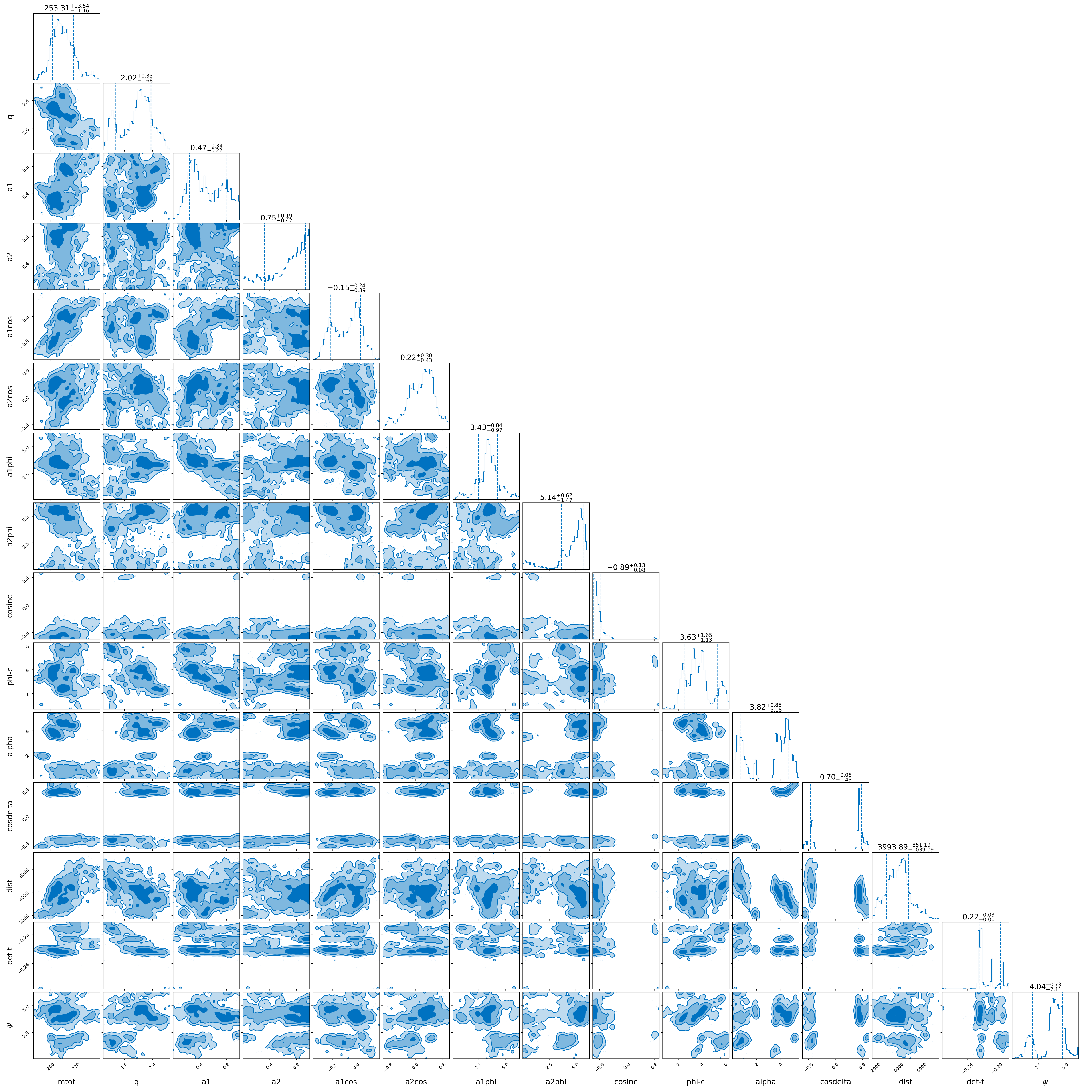}
    \caption{Parameter estimation of the BBH scenario for the GW190521. The parameters are the total mass in the detector frame~[$M_{\odot}$], mass ratio$\geq1$, magnitudes and angular orientations of the dimensionless spins, cosine of the orbital inclination, coalescence phase, right ascension, cosine of declination, luminosity distance~[Mpc], reference detection time~[s] and polarization angle. Angular quantities are in radians and definitions of the spin orientations are according to the assumed coordinate system of the \texttt{gwsurrogate} package~\cite{2014PhRvX...4c1006F}.}
    \label{fig:190521BBHPE}
\end{figure*}

\begin{figure}
    \centering
    \includegraphics[width=\columnwidth]{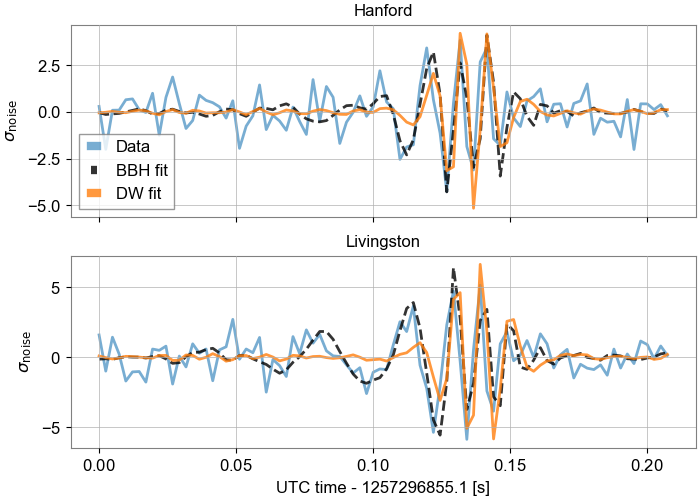}
    \caption{Whitened best-fitting TDM and BBH signals for the data of GW191109\_010717. The y axis is in units of the standard deviation of the noise.}
    \label{fig:191109}
\end{figure}

Let us now elaborate on our findings for GW190521. Comparing the evidence for the TDM hypothesis against the BBH hypothesis, our analysis yields a log-evidence difference of $\Delta \ln \mathcal{B}\approx 21$ favoring the BBH scenario. In Fig. \ref{fig:190521} we show the best-fitting BBH and TDM templates to the data of Hanford and Livingston. The log-likelihood difference for these two cases is about $\Delta \ln \mathcal{L}\approx25$ favoring the BBH fit. The parameter estimations for the TDM and BBH scenarios are shown in Figs. \ref{fig:190521PE} and \ref{fig:190521BBHPE} respectively. Although Fig. \ref{fig:190521} shows similar fits for both hypotheses, there is a small phase evolution difference in favor of the BBH fit. The similar fits are mainly generated by the $m_0$ interaction. The estimated $g_{\rm DW}$ values are orders of magnitude lower than its sensitivities, implying that it has a negligible contribution. This is not surprising as the two couplings generate very different signals. In Fig. \ref{fig:190521PE} the TDM fits have a very precise estimation for the orientation of the DW. This is due to the speed of the DW, $v_{\rm DW}$, being $\mathcal{O}(10^2-10^3)$ times smaller than the GW speed. Therefore, the necessary time delay of the signals between the two detectors can be achieved only within a small parameter volume, which could be more easily achieved with a propagation speed close to the speed of light. Similar arguments hold true as well for the second-to-best fit, i.e., the event GW191109\_010717, which, for comparison, we show in Fig. \ref{fig:191109}.

Despite the BBH hypothesis being favored throughout the analyzed events, the similarity of the fitted BBH and TDM signals raises the question of whether a TDM signature could potentially be captured by a BBH template fit. This would imply that BBH searches could, in principle, catch TDM signals in future runs. 
To determine whether the latter is technically feasible, we fit BBH templates to the best-fitting TDM template for GW190521. We find that the SNR of the best-fitting BBH template to the TDM signal is about $
\sim$90\% of the actual SNR of the simulated TDM signal in the detector. This implies that template-based BBH searches for events in LIGO are potentially triggered by TDM signatures in detector data. 

\subsection{TDM as glitches}\label{glitches}
The LIGO detectors are frequently plagued by glitches, which are reoccurring signatures in the data that cannot be assigned to astrophysical sources. Despite their failure to match merger events, glitches play an important role in the data analysis challenges often creating false detection alarms. Currently, than 20 distinct glitch types appearing in LIGO data have been classified and large efforts are being made to investigate their origin. While the sources of a few glitches have been identified, some are still not fully understood. Here we investigate whether glitches can limit TDM searches in the GW detector data. We consider two cases. First we examine the possible effects of the ``blip'' glitches, and then we analyze a fast scattering glitch around the 200214\_224526 trigger.

\subsubsection{Blip glitches}
\begin{figure}
   \includegraphics[width=\columnwidth]{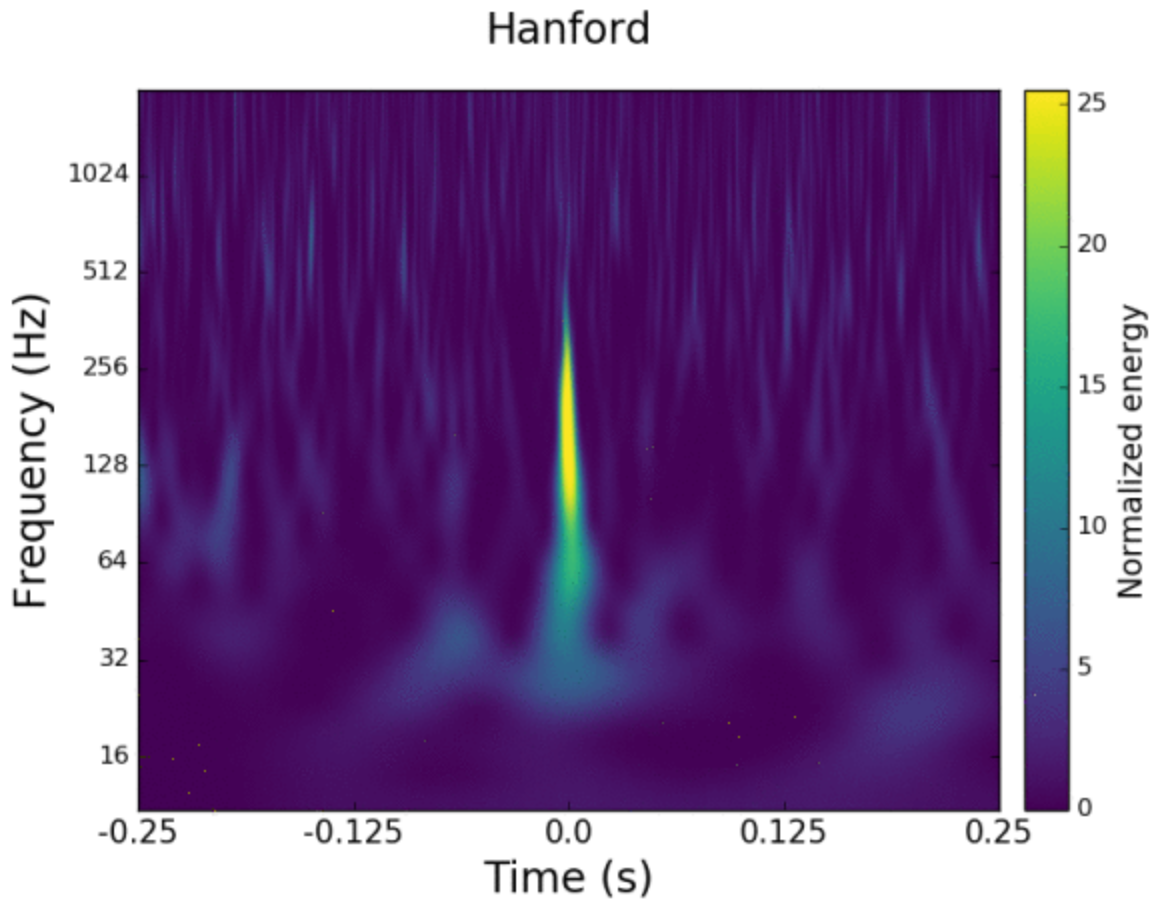}

\includegraphics[width=\columnwidth]{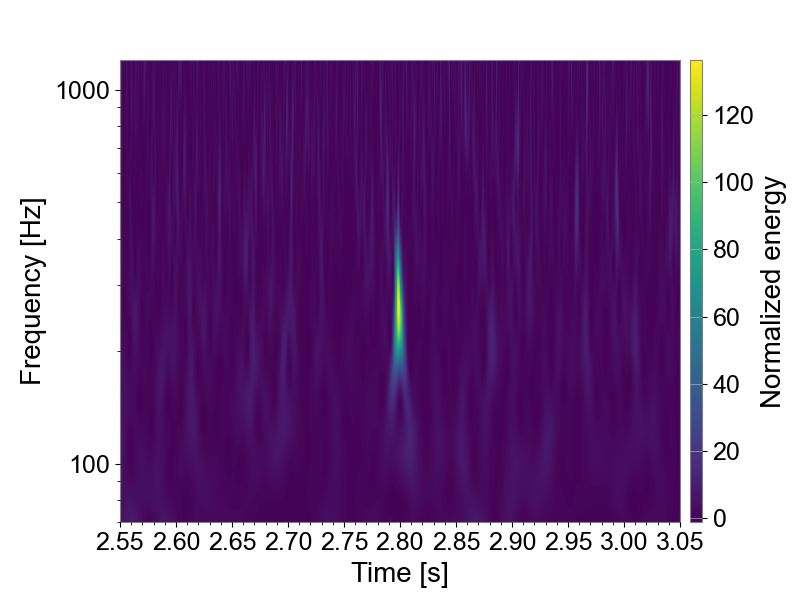}   
     \caption{Q-transforms of (top) an exemplary blip recorded in the Hanford detector (from the GravitySpy collection \cite{Zooniverse,Zevin_2017}) and (bottom) a simulated TDM signal.}
       \label{fig:Ng2}
\end{figure}

The so-called ``blips'' are identified by their characteristic shape in the time-frequency plane. TDM signals can be very similar to blips if the DW traverses an interferometer arm perpendicularly, creating a very short signal. In Fig. \ref{fig:Ng2} we provide concrete proof of this claim. We display an instance of a blip that appeared in LIGO data next to a generic (short) TDM signal demonstrating their clear similarity. The top panel shows a piece of actual data from the Hanford detector in which a blip appears, whereas the bottom displays the q-transform for a TDM within some noisy background with a sufficiently high SNR. Although numerous glitches remain unassigned to a particular phenomenology, our analysis reveals that blip-type glitches, on average, are the glitches most likely to produce data signatures sufficiently similar to TDM.

Blips appear in LIGO data at a rate of about $\mathcal{O}(10)$ per hour \cite{Cabero_2019}. This rate is reduced when filtering for events that correlate over both LIGO detectors with only a few fractions of a second delay. With a resulting order of $\mathcal{O}(10)$ per year-long observing run, correlated blips embody a systematic effect suitable for the application of TDM searches. Note also that, in comparison to the TDM fit to GW signals, the signal correlations are less tightly constrained in the time domain due to the absence of the imposed GW velocity. Concretely, we include blips with time delays between the detectors that are 2-3 orders of magnitude larger than the GW searches. This opens up a new, previously ``untouched'' regime in TDM parameter space and constitutes a huge advantage with respect to the previous GW-TDM fitting. In practice, to investigate the role of blips in TDM searches, we compare multiple realizations of blips with signatures produced by TDM in a realistic detector setup. To avoid searching for blips in real data, we instead use randomly generated glitch data created by a machine-learning-powered glitch generator \cite{Powell_2023}. The latter tool is specifically trained to provide realistic glitch data for LIGO detectors and thus establishes a good alternative for actual glitch data.

For the blip analysis, we recycle the setup used for the GW data fit. Our simulation pipeline infuses the glitch into the detector data at a given time $t_0$. We then simulate the DW of TDM passing through the detector such that the signal appears at $t_0$ as well. Again, we apply a similar sampling scheme to find the best fit for the model parameters $m_{\text{dark}}$, $m_0$, $g_{\rm DW}$, $v$, $N_A/N_\phi$, $\theta$, and $\phi$, where the latter angles parametrize the projection of the DWs unit normal vector $\mathbf{n}$ onto the detector plane. Prior to the evaluation of the likelihood, glitch and TDM signals are whitened. We do not perform a joint analysis of glitches correlated between LIGO detectors as there exist readily available instances of correlated blips.\footnote{Again, correlated blips are known to exist, but unlike the GW events, they have to be searched for in LIGO data ``by hand'' which goes beyond the frame of this work.}

\begin{figure}
       \includegraphics[width=\columnwidth]{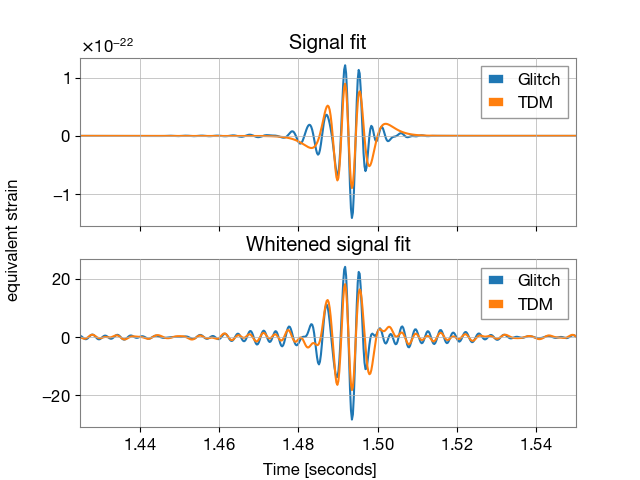}

       \includegraphics[width=\columnwidth]{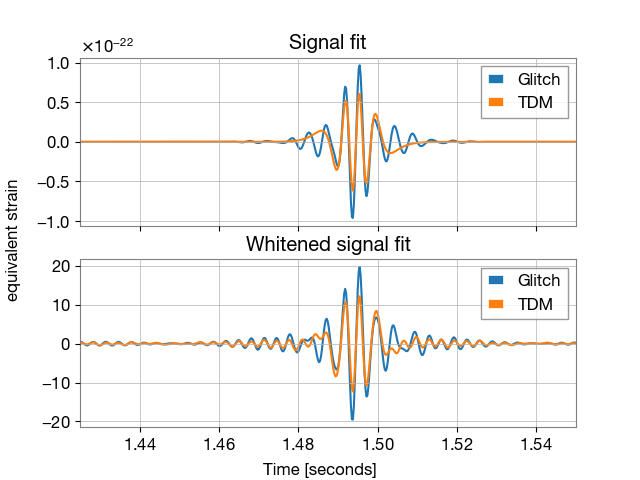}

     \caption{Fits for simulated TDM signatures inside Livingston onto blip glitches generated by the GlitchGen AI tool \cite{Powell_2023}. The upper and lower blips are independently generated. The y axes of the whitened signals are in units of the standard deviation of the noise.}
       \label{fig:Fits}
\end{figure}

We display two exemplary best fits in Fig. \ref{fig:Fits}. The simulations displayed here result from multiple runs of the sampling scheme where, after each run, we adapt the priors to narrow down the desired parameter domain in each step. We find that the glitches can well resemble TDM signatures for specific parameter regions. Concretely, we find that around $m_\text{dark}\sim 2\cdot 10^{-11}$ eV, $m_0 \sim 10^{-11}$ eV, $N_A/N_\phi = 2 $, and $v\sim 10^{-2}\cdot c$ TDM signals most likely mimic the signature of blip glitches. The coupling constant related to the test mass acceleration, $g_\text{DW}$, is effectively set to zero by the fitting procedure. This matches our expectation, as the signatures caused by the test mass acceleration, depicted in Fig. \ref{fig:sig_examples}, are highly dissimilar to oscillation observed for TDM DW crossings in interferometers. This means that the blip glitches would not be a big problem for the searches looking for the signals sourced by $g_{\rm DW}$ as they can be distinguished. Generally, for all fit parameters used in the glitch analysis, the obtained fit values are astrophysically relevant as they satisfy the regularity conditions calculated in Sec. \ref{intro} regarding the DW crossings per unit time.
Finally, it is essential to emphasize that the majority of blip glitches exhibit no correlation between the detectors, indicating that TDM alone cannot account for all blips observed in LIGO data. Instead, the matches identified should be interpreted as a cautionary indication that potential TDM signatures may be mistakenly dismissed as glitches.

\begin{figure}
    \centering
    \includegraphics[width=0.48\textwidth]{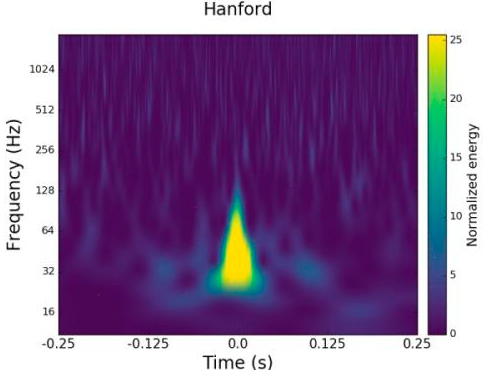}
    \caption{Example of a Tomte-type glitch (from the GravitySpy collection \cite{Zooniverse,Zevin_2017})}
    \label{fig:TomteGlitch}
\end{figure}

So far, we only considered blips fitted to DW signals but as mentioned above, there is an abundance of different types of glitches present in LIGO data. The selection of potential fit candidates from this ensemble of glitches is based on their shape in frequency space. Besides blips, the so-called tomte glitches provide a promising signature similar those of TDM. An instance of a tomte glitch is displayed in Fig. \ref{fig:TomteGlitch}. While there are similarities regarding the general shape of the signal, the power in the lower frequency regime is spread out significantly more than what is achievable with TDM. After careful investigations of the fit parameter space, we conclude that, in fact, the dissimilarities between tomte glitches and TDM signals are too large to provide a viable explanation of these glitches in terms of TDM.

\subsubsection{Fast scattering glitch}
\begin{figure*}
    \centering
    \includegraphics[width=\textwidth]{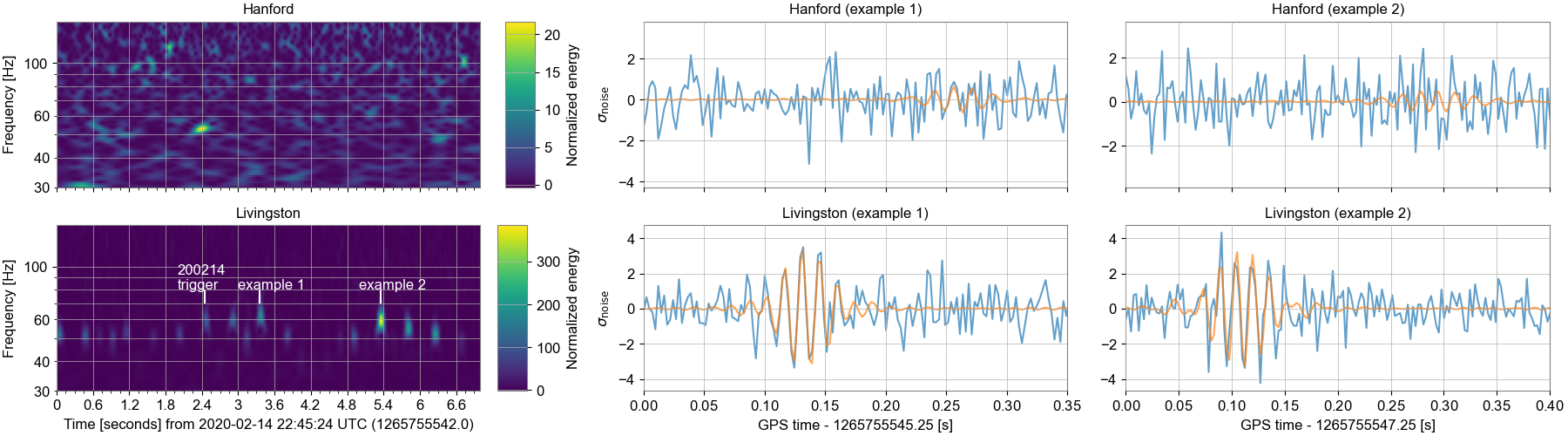}
    \caption{Left: Q-transforms of the (top row) Hanford and (bottom row) Livingston data around the 200214\_224526 trigger. Middle and Right: the detector data (blue) and the TDM signal fit (orange) for the most powerful bursts of the glitch (pointed in the left bottom panel)}
    \label{fig:200214}
\end{figure*}

While analyzing the BBH events, we also examine the marginal triggers indicative of intermediate-mass black holes. Among these, we observe that the fast scattering glitch in Livingston, which triggers the event labeled 200214\_224526, exhibits features potentially resembling those of TDM. Typically, the fast scattering glitch manifests as repeated bursts (see Fig. \ref{fig:200214}). Notably, the strain data of individual bursts demonstrates strong resemblances to TDM signatures, prompting us to apply a fitting scheme to event 200214\_224526 as well. In this specific trigger, we focus on analyzing the two most powerful bursts. Employing the same analysis approach as for the BBH matches outlined in Sec. \ref{bbhs}, we jointly consider data from both detectors. Considering the significantly different signal shape generated by the test mass acceleration, we omit that effect here, i.e., we only consider the signals produced by the photon mass term interaction with the $m_0$ parameter in Eq. \eqref{equ:4}. Figure \ref{fig:200214} illustrates the strain data from the detectors alongside the fitted TDM templates. The striking resemblance between the examined bursts and TDM underscores the significance of thorough considerations of TDM as potential glitch triggers.

Summarizing our analysis, we provide concrete evidence suggesting that DW composed of TDM can be explored using GW interferometers. TDM signatures exhibit notable compatibility with BBH events, and BBH template searches possess the potential to detect putative TDM signals within interferometer data. Additionally, we demonstrate the presence of recurring glitches within LIGO data exhibiting a time-frequency profile akin to that of TDM signals, albeit within a restricted parameter space. These fits exhibit resilience against noise and are anticipated to occur with sufficient frequency to facilitate correlation testing between both LIGO detectors. The investigation of such correlations is deferred to future analyses involving actual detector data.

\section{Discussion}\label{discus}
In this article, we examined the detectability of TDM using GW interferometers and performed a preliminary analysis of LIGO data in search of TDM signals. First, our findings highlight that despite having a distinct physical origin and fewer parameters in the modeling compared to CBC sources (8 vs. 15), signals from TDM can exhibit strong similarities to BBH mergers. In particular, our analysis revealed that the GW190521 event could also be accounted for by the TDM hypothesis, although it is not preferred over the BBH explanation in a direct comparison. 
Additionally, we observed that CBC searches have the capability to capture TDM signals. In our analysis, a BBH template managed to generate approximately 90\% of the actual SNR of the best-fitting TDM template for GW190521. However, as this represents just one instance, the detectability of TDM signals with varying parameters may prove to be more or less feasible.

Second, we identified resemblances between TDM signals and glitches present in the data. Specifically, we observed that blip glitches exhibit characteristics similar to a TDM signal within a specific parameter range ($m_\text{dark}\sim 2\cdot 10^{-11}$ eV, $m_0 \sim 10^{-11}$ eV, $N_A/N_\phi = 2 $, $v\sim 10^{-2}\cdot c$, $g_\text{DW}\rightarrow0$), suggesting that the identification of a TDM signal with these parameters could potentially be obscured by the presence of blip glitches in measurement data. Additionally, we noted that various other types of glitches could also pose potential challenges for TDM detections. Most notably, the TDM model successfully fitted a fast scattering glitch. In combination, these findings highlight potential systematic issues for TDM searches in GW data.

Based on the insights gleaned from this analysis, we can devise novel strategies aimed at facilitating the potential identification of TDM signatures in data collected during future GW detection runs. All proposed modifications should be regarded as supplements to the existing analysis pipeline and are deemed technically feasible: the most straightforward modification to GW data analysis involves an adjustment of the detection delay time window. In light of the slower speeds of DWs, we propose an extension of the time window between detectors currently tailored to accommodate the fast ($\approx c$) propagation speed of GWs. Given that the expected speeds of TDM DWs could be as low as $10^{-3}c$, significantly longer time windows would be necessary for effective searches. Note, however, that this adjustment is anticipated to yield a less potent GW search due to the increased number of expected noise triggers within the extended time window.
The second modification we propose to enhance the identification of TDM in detector data involves removing assumptions regarding GW polarizations from the analysis pipelines. Both the CBC and unmodeled burst pipelines currently incorporate a signal transformation between detectors based on the assumed polarization of the GWs. We recommend abandoning or modifying this transformation to account for the two possible and independent effects of TDM: (i) modifications of the phase velocity of light, as described by Eq. \eqref{eq:phase_vel}, and (ii) changes in the fine-structure constant, $\alpha$, as exemplified by Eq. \eqref{equ:beam_mirror}.
Last, the antenna patterns of the detectors should be adjusted to accommodate the narrow physical thickness of the DWs in comparison to the GW wavelength, which cannot be assumed to be in the long ``wavelength" limit. Antenna patterns should also be adapted to incorporate the distinct mechanisms through which TDM signals are generated within the interferometer. The relevant phase shifts can be induced by alterations in the fine-structure constant or changes in the phase velocity of light, which structurally differ from the effect of GWs on test masses. Unlike GW effects, which are equivalent in different reference frames, these TDM-induced phase shifts exhibit unique characteristics that necessitate tailored adjustments to antenna patterns.

In combination, the aforementioned guidelines provide a framework for facilitating potential detections of TDM with GW interferometers, a prospect demonstrated to be well within technical feasibility in this study. The pursuit of DM holds significant promise for advancing our comprehension of its elusive nature, underscoring the imperative to explore every viable avenue in this endeavor. Our investigation serves as a proof of concept intended to catalyze future endeavors focused on the search for TDM using actual detector data.

\section*{Acknowledgments}
The authors thank Robert Brandenberger, Gayathri V., Zsuzsa M\'arka, Szabolcs M\'arka, Imre Bartos and Juan Calderon Bustillo for discussions and feedback. This document was reviewed by the LIGO Scientific Collaboration under the document number P2300297. L.H. is supported by funding from the European Research Council (ERC) under the European
Unions Horizon 2020 research and innovation programme grant agreement No 801781 and
by the Swiss National Science Foundation grant 179740. L.H. further acknowledges support
from the Deutsche Forschungsgemeinschaft (DFG, German Research Foundation) under
Germany's Excellence Strategy EXC 2181/1 - 390900948 (the Heidelberg STRUCTURES
Excellence Cluster).

This research has made use of data or software obtained from the Gravitational Wave Open Science Center (gwosc.org), a service of the LIGO Scientific Collaboration, the Virgo Collaboration, and KAGRA. This material is based upon work supported by NSF's LIGO Laboratory which is a major facility fully funded by the National Science Foundation, as well as the Science and Technology Facilities Council (STFC) of the United Kingdom, the Max-Planck-Society (MPS), and the State of Niedersachsen/Germany for support of the construction of Advanced LIGO and construction and operation of the GEO600 detector. Additional support for Advanced LIGO was provided by the Australian Research Council. Virgo is funded, through the European Gravitational Observatory (EGO), by the French Centre National de Recherche Scientifique (CNRS), the Italian Istituto Nazionale di Fisica Nucleare (INFN) and the Dutch Nikhef, with contributions by institutions from Belgium, Germany, Greece, Hungary, Ireland, Japan, Monaco, Poland, Portugal, Spain. KAGRA is supported by Ministry of Education, Culture, Sports, Science and Technology (MEXT), Japan Society for the Promotion of Science (JSPS) in Japan; National Research Foundation (NRF) and Ministry of Science and ICT (MSIT) in Korea; Academia Sinica (AS) and National Science and Technology Council (NSTC) in Taiwan.

\bibliographystyle{apsrev4-2}
\bibliography{references}

\end{document}